\documentclass[showpacs,twocolumn,superscriptaddress]{revtex4-1}
\usepackage{doi}
\usepackage{hyperref}
\hypersetup{
  colorlinks=true,        
  linkcolor=blue,         
  citecolor=cyan,         
}

\usepackage{graphicx}
\usepackage{dcolumn}
\usepackage{bm}
\usepackage{color}
\usepackage{enumitem}
\usepackage{amsmath}
\usepackage{amssymb}
\usepackage{orcidlink}
\usepackage{graphicx} 
\usepackage{subcaption}

\begin{document}

\title{Gravitational waveforms from periodic orbits around a dyonic ModMax black hole}

\author{Mirzabek Alloqulov,\orcidlink{0000-0001-5337-7117}}
\email{malloqulov@gmail.com}

\affiliation{School of Physics, Harbin Institute of Technology, Harbin 150001, People’s Republic of China}
\affiliation{University of Tashkent for Applied Sciences, Str. Gavhar 1, Tashkent 100149, Uzbekistan}

\author{Sanjar Shaymatov, \orcidlink{0000-0002-5229-7657}}
\email{sanjar@astrin.uz}
\affiliation{Institute of Fundamental and Applied Research, National Research University TIIAME, Kori Niyoziy 39, Tashkent 100000, Uzbekistan}
\affiliation{Institute for Theoretical Physics and Cosmology,
Zhejiang University of Technology, Hangzhou 310023, China}

\affiliation{Tashkent State Technical University, 100095 Tashkent, Uzbekistan}

\author{Bobomurat Ahmedov,\orcidlink{0000-0002-1232-610X}}
\email{ahmedov@astrin.uz}

\affiliation{Institute for Advanced Studies, New Uzbekistan University, Movarounnahr str. 1, Tashkent 100000, Uzbekistan}

\affiliation{School of Physics, Harbin Institute of Technology, Harbin 150001, People’s Republic of China}

\affiliation{Institute of Theoretical Physics, National University of Uzbekistan, Tashkent 100174, Uzbekistan}

\author{Tao Zhu, \orcidlink{0000-0003-2286-9009}}
\email{zhut05@zjut.edu.cn}
\affiliation{Institute for Theoretical Physics and Cosmology, Zhejiang University of Technology, Hangzhou 310023, China}
\affiliation{United Center for Gravitational Wave Physics (UCGWP), Zhejiang University of Technology, Hangzhou, 310032, China}

\begin{abstract}

In this work, we study the gravitational waveforms from the periodic orbits of a massive particle around a dyonic ModMax black hole. We begin with a brief analysis of the spacetime and then examine how its parameters influence the dynamics of a massive neutral particle using the Lagrangian formalism. In particular, we compute the characteristics of marginally bound orbits and innermost stable circular orbits. Our results show that the values of these quantities increase with the black hole charge $Q$ and the screening parameter $\gamma$. We then plot various periodic orbits, characterized by the integers ($z$,$w$,$v$). Finally, we present the gravitational waveforms associated with extreme mass ratio inspirals, consisting of a stellar-mass compact object orbiting a supermassive black hole.

\end{abstract}
\maketitle

\section{Introduction}

Black holes (BHs) formed at the end state of evolution of massive stars remain among the most intriguing predictions of general relativity (GR). Compact BHs serve as cosmic laboratories for testing gravity theories in the strong-field regime. GR, by describing gravity as the curvature of spacetime \cite{Einstein1916}, revolutionized our understanding of the nature of the universe.  The different observations, performed in both weak (Solar System) and strong field (BHs environment) regimes, have validated GR. For instance, the groundbreaking observations of the BH shadows in M87$^*$~\cite{Akiyama19L1, Akiyama19L6} and SgrA$^*$~\cite{Event}, together with the first direct detections of gravitational waves (GWs)~\cite{Abbott_2016} originating from BH and neutron star mergers, have marked major milestones in modern astrophysics. 

GR, despite its great success, has limitations, as it is a classical theory. These include the presence of physical singularities arising in cosmological solutions and in astrophysics from gravitational collapse, and their inconsistency with quantum field theory. Consequently, modified~\cite{Makarenko_2012, Nojiri_2017, Shankaranarayanan_2022} and alternative theories of gravity~\cite{Kleihaus2023, Ashtekar_2021} have been introduced to resolve these spacetime singularities and pave the way for a theory of quantum gravity. Among them, the Modified Maxwell (ModMax) theory plays an important role.

The dyonic ModMax BH represents an exact solution in GR coupled to nonlinear electrodynamics, in the framework of ModMax theory. This theory, introduced as a nonlinear extension of Maxwell's equations that preserves both electromagnetic duality invariance and conformal symmetry, allows for BH solutions characterized by both electric and magnetic charges; consequently, the term ``dyonic'' was introduced. The ModMax Lagrangian depends on a nonlinearity parameter \(\gamma\), which modulates the strength of electromagnetic interactions and leads to screening effects on the charges. In Einstein-ModMax gravity, the static spherically symmetric dyonic BH space-time metric resembles the Reissner-Nordström solution~\cite{Reissner1916, Nordstrom1918} but with modifications due to the screening parameter \(\gamma\), which affects properties such as the horizon structure, thermodynamics, and optical phenomena such as shadows and gravitational lensing. Recent studies have explored its geodesic motion, accretion processes, quasinormal modes, and gravitational lensing, revealing potential observational signatures that could distinguish it from standard charged BHs (see, e.g., \citep{Bandos2020, Flores-Alfonso2020, Narzilloev20a, Pantig2022, Shaymatov22a,Dengiz2025, Khosravani:2024xvz,Shahzad:2024ljt, HabibMazharimousavi:2022ppz}).

Future space-based detectors, such as the Laser Interferometer Space Antenna (LISA) \cite{Amaro-Seoane2017} and the Chinese Taiji mission \cite{HuTaiji}, aim to explore lower frequency gravitational-wave signals, including extreme mass-ratio inspirals (EMRIs) in which a stellar-mass object orbits a supermassive BH (SMBH) \cite{Hughes01EMRI,Amaro-Seoane18LRR,Babak17PRD}. In general, the trajectory of a stellar-mass object orbiting the SMBH is characterized by the tightly bound orbits during its inspiral phase and can be well approximated by periodic orbits. The periodic orbits are a special type of orbit of a test particle that can return to the same position and velocity (or direction) relative to the BH after a finite time. This kind of orbits often exhibit zoom-whirl behavior and have been analyzed thoroughly in Schwarzschild and Kerr spacetimes by using a characterization scheme described by three topological numbers \cite{Levin:2008ci, Levin:2009sk, Bambhaniya:2020zno, Rana:2019bsn, Levin:2008mq}. By using the same taxonomy, a large amount of research has been carried out on periodic orbits within various BH spacetimes, see refs.~\cite{Babar:2017gsg, Liu:2018vea, Lin:2023rmo, Yao:2023ziq, Lin:2022llz, Chan:2025ocy, Wang:2022tfo, Lin:2023eyd, Haroon:2025rzx, Habibina:2022ztd, Zhang:2022psr, Lin:2022wda, Gao:2021arw, Lin:2021noq, Deng:2020yfm, Tu:2023xab, Zhou:2020zys, Gao:2020wjz, Deng:2020hxw, Azreg-Ainou:2020bfl, Wei:2019zdf, Pugliese:2013xfa,Zhang:2022zox, Healy:2009zm, Wang:2025wob, Alloqulov:2025bxh, Wei:2025qlh} and references therein. The gravitational waveforms from the peroidic orbits of an EMRIs have also been studied in various BHs \cite{Zhang:2025wni, Tu:2023xab, Yang:2024lmj, Shabbir:2025kqh, Junior:2024tmi,  Jiang:2024cpe, Yang:2024cnd, Li:2024tld, QiQi:2024dwc, Haroon:2025rzx, Alloqulov:2025ucf, Wang:2025hla, Lu:2025cxx, Zare:2025aek, Gong:2025mne, Li:2025sfe, Choudhury:2025qsh, Chen:2025aqh, Deng:2025wzz, Li:2025eln, Zahra:2025tdo, Sharipov:2025yfw, Ahmed:2025azu, Glampedakis2002}.

In this work, we investigate the dynamics of periodic “zoom-whirl” orbits around a static, charged dyonic ModMax BH and the waveforms of the gravitational wave radiation from certain periodic orbits. 
The paper is organized as follows. In Section \ref{Spacetime}, the spacetime of dyonic ModMax BH is discussed, and geodesic orbits in its close environment are explored. In the next Section~\ref{periodic}, we investigate the periodic orbits around the dyonic ModMax BH. We compute key orbital thresholds, including the marginally bound orbit (MBO) and the ISCO radii, and demonstrate how the ModMax screening parameter $\gamma$ (and the associated charge $Q$) modifies these features compared to the Schwarzschild and Reissner–Nordström black-hole cases. In Section~\ref{numerical}, the gravitational waveforms of periodic orbits around the dyonic ModMax BH are numerically calculated. We model the gravitational waveforms generated by EMRIs in this spacetime, using the numerical kludge approach \cite{Poisson2014}. These waveforms encode distinct signatures of periodic orbits, offering a potential observational probe of deviations from the Einstein GR in strong-field regimes. The case of the extreme mass-ratio inspiral (EMRI) system is studied, where the EMRI system involves a stellar-mass object moving along the periodic orbit around a supermassive dyonic ModMax BH. Section~\ref{conclusion} summarizes our main findings.

\section{Spacetime of dyonic ModMax BH}
\label{Spacetime}

\begin{figure*}[!htb]
\includegraphics[scale=0.4]{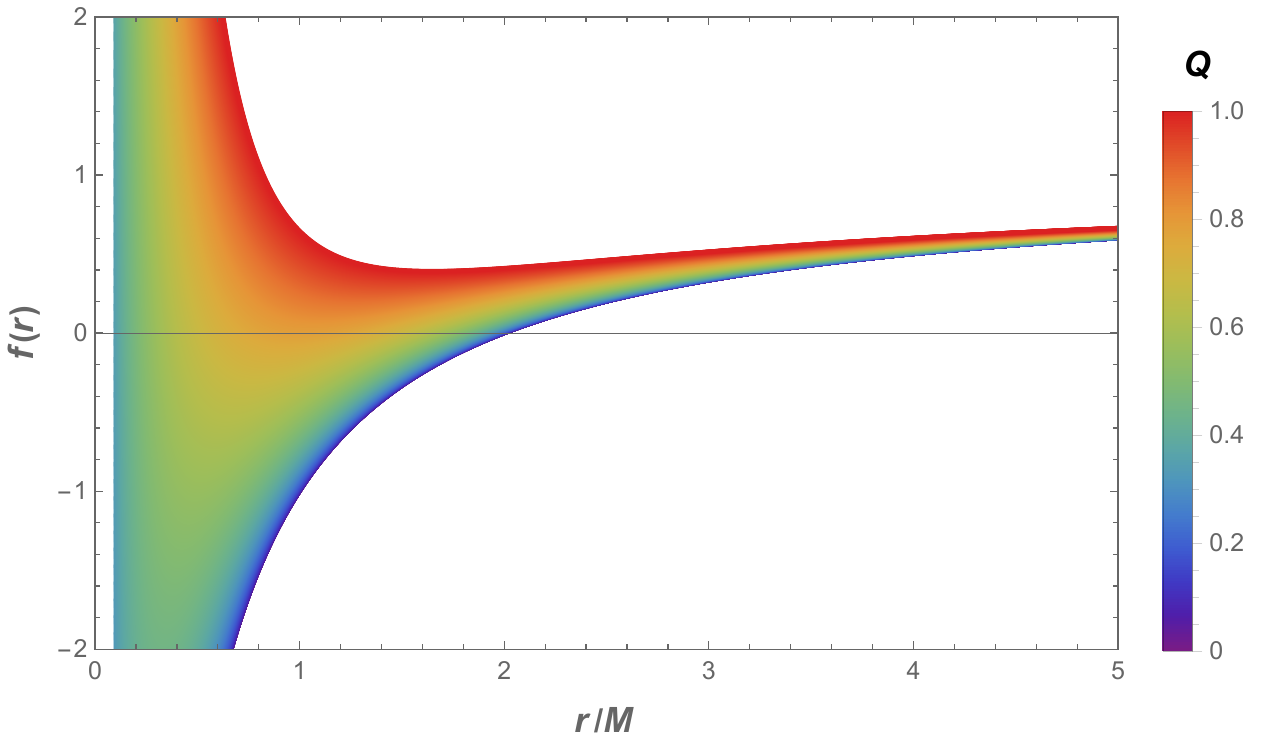}
\includegraphics[scale=0.4]{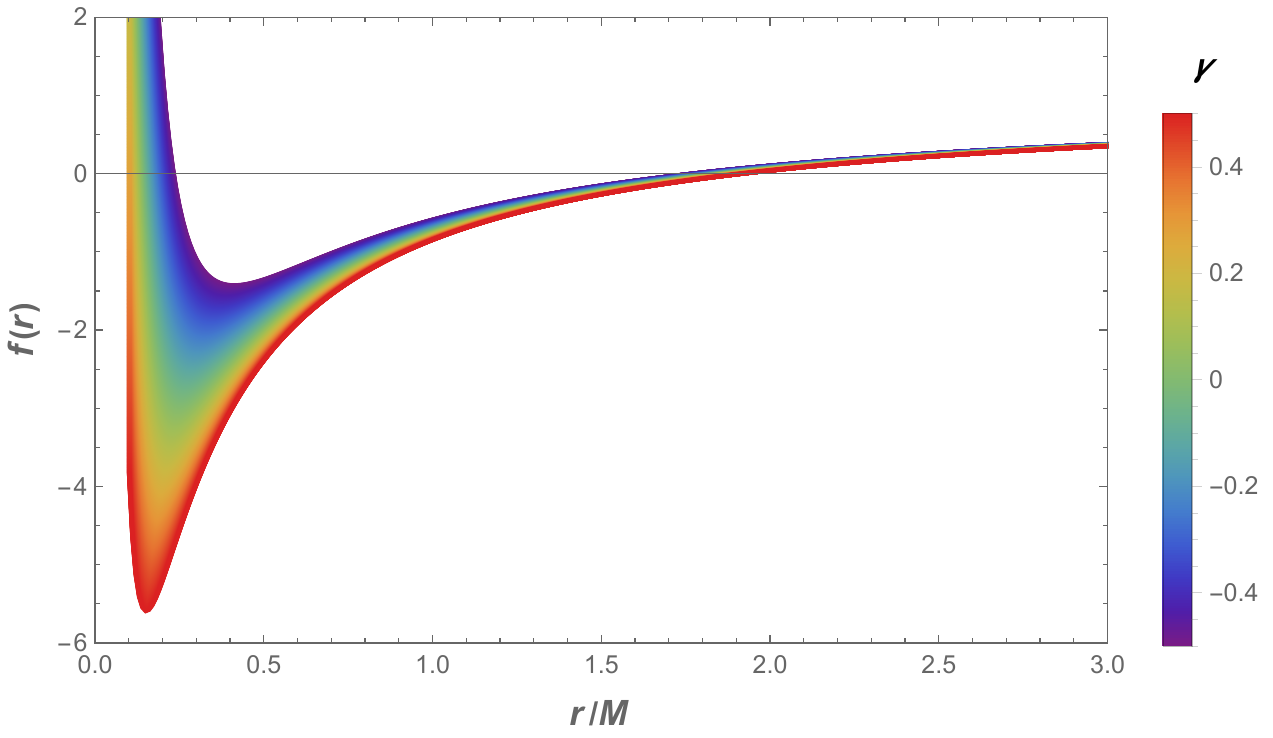}
\caption{The left panel shows the radial dependence of the metric function $f(r)$ for different values of the BH charge. The screening parameter is fixed as $\gamma=-0.5$. The right panel illustrates the radial dependence of the metric function $f(r)$ for different values of the screening parameter $\gamma$. The BH charge is fixed as $Q=0.5$.}
\label{fig:f(r)}
\end{figure*}
\begin{figure}[!htb]
    \centering
   \includegraphics[scale=0.4]{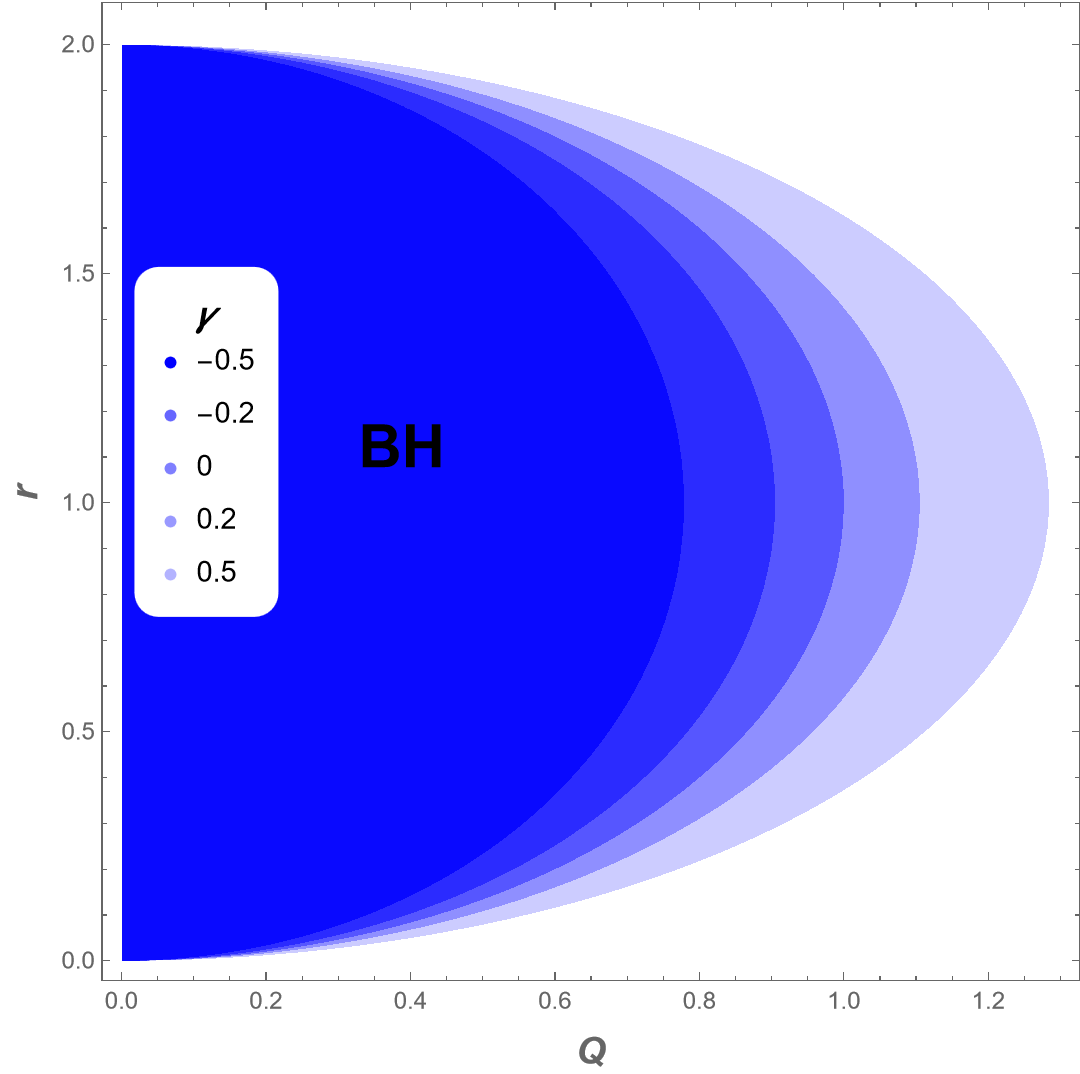}
    \caption{The phase diagram demonstrates the existence of the ModMax BH in the $(Q,r)$ plane. The blue region corresponds to the slice of the parameter space where $f(r,Q)\leq0$, indicating the presence of the ModMax BH. The different opacities of the blue refer to the different values of the screening parameter $\gamma$.}
    \label{fig:phasediagram}
\end{figure}
\begin{figure*}[!htb]
\includegraphics[scale=0.4]{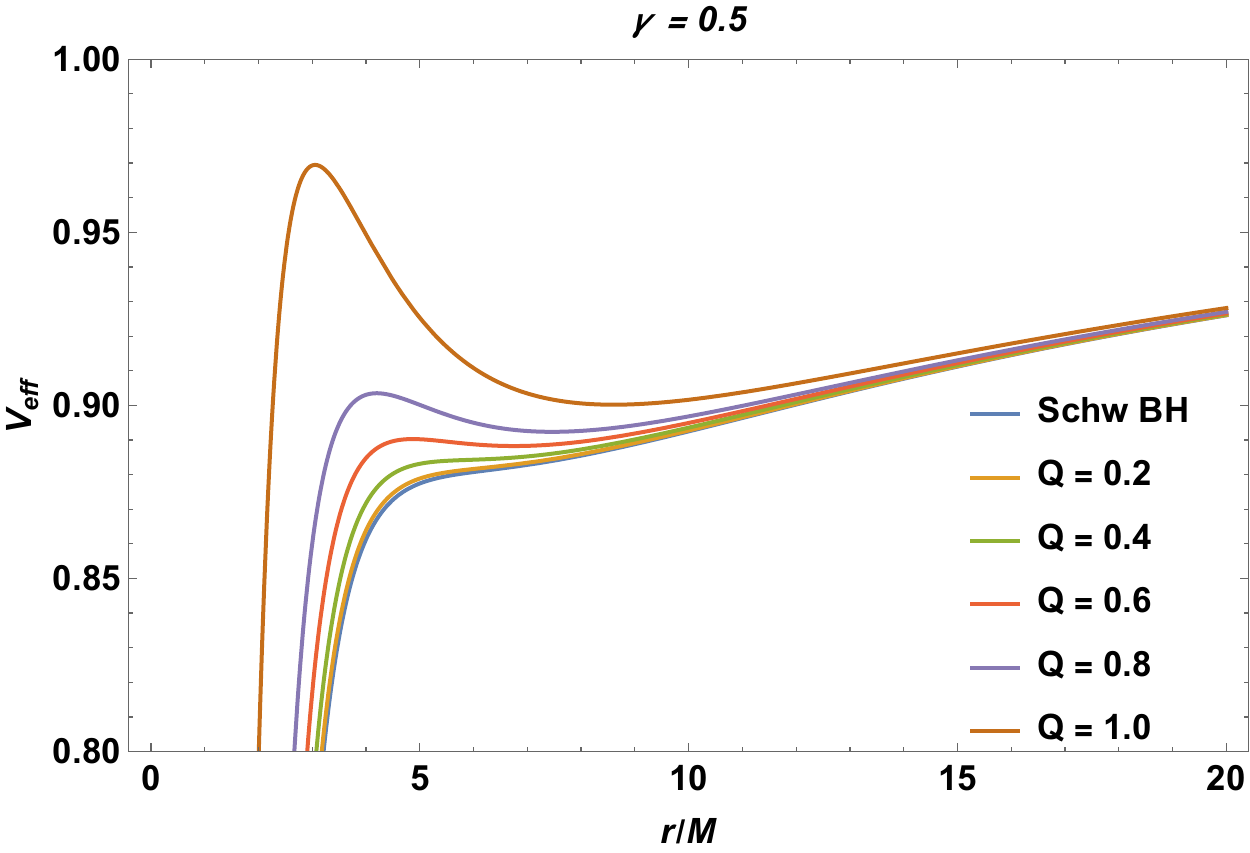}
\includegraphics[scale=0.4]{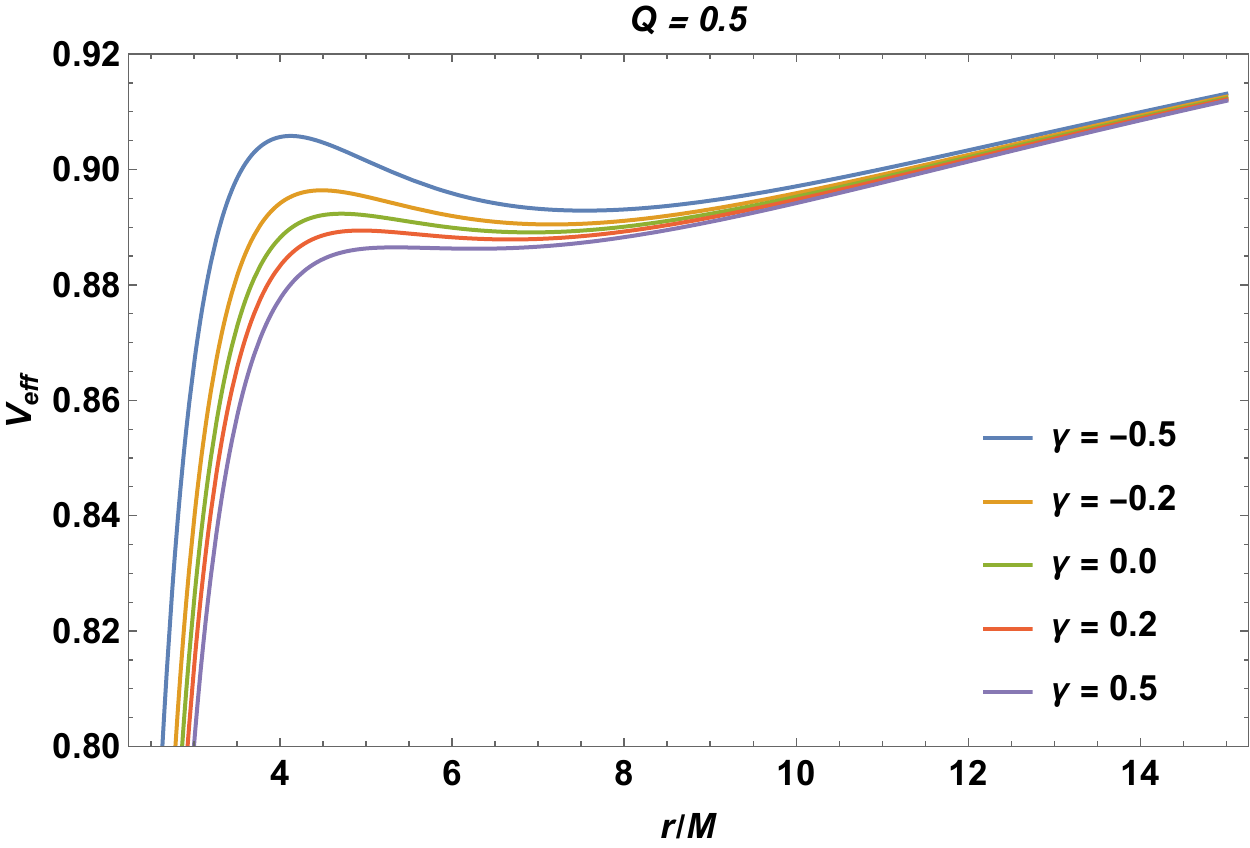}
\includegraphics[scale=0.4]{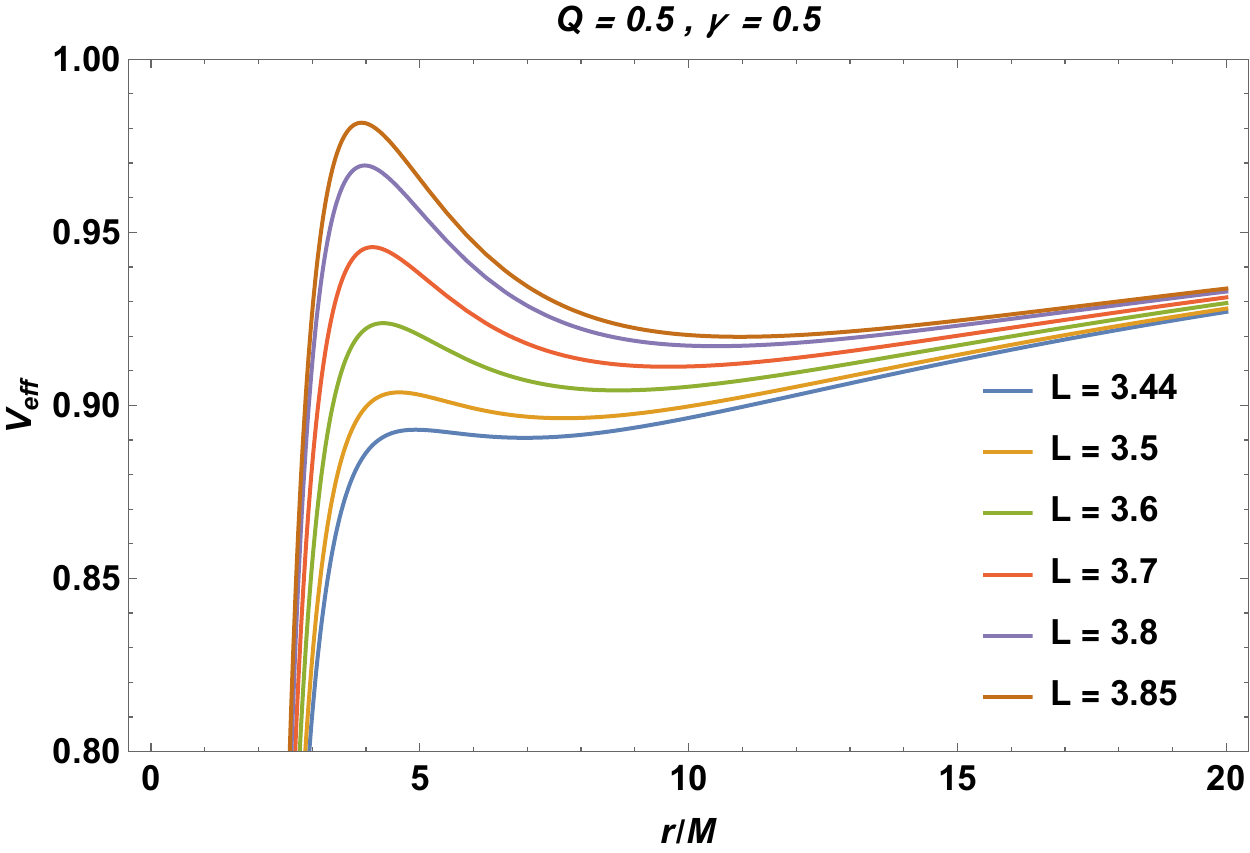}
\caption{The top-left panel shows the radial dependence of the effective potential of the test particles around the ModMax BH for different values of the BH charge $Q$. Here, we set the screening factor as $\gamma=0.5$. The top-right panel illustrates the radial dependence of the effective potential for different values of the screening factor $\gamma$. Here, the BH charge is equal to $0.5$. Bottom panel: The plot demonstrates the radial dependence of the effective potential for different values of the orbital angular momentum. The other parameters are fixed as $Q=0.5$ and $\gamma=0.5$.}
\label{fig:eff}
\end{figure*}
\begin{figure*}[!htb]
\includegraphics[scale=0.4]{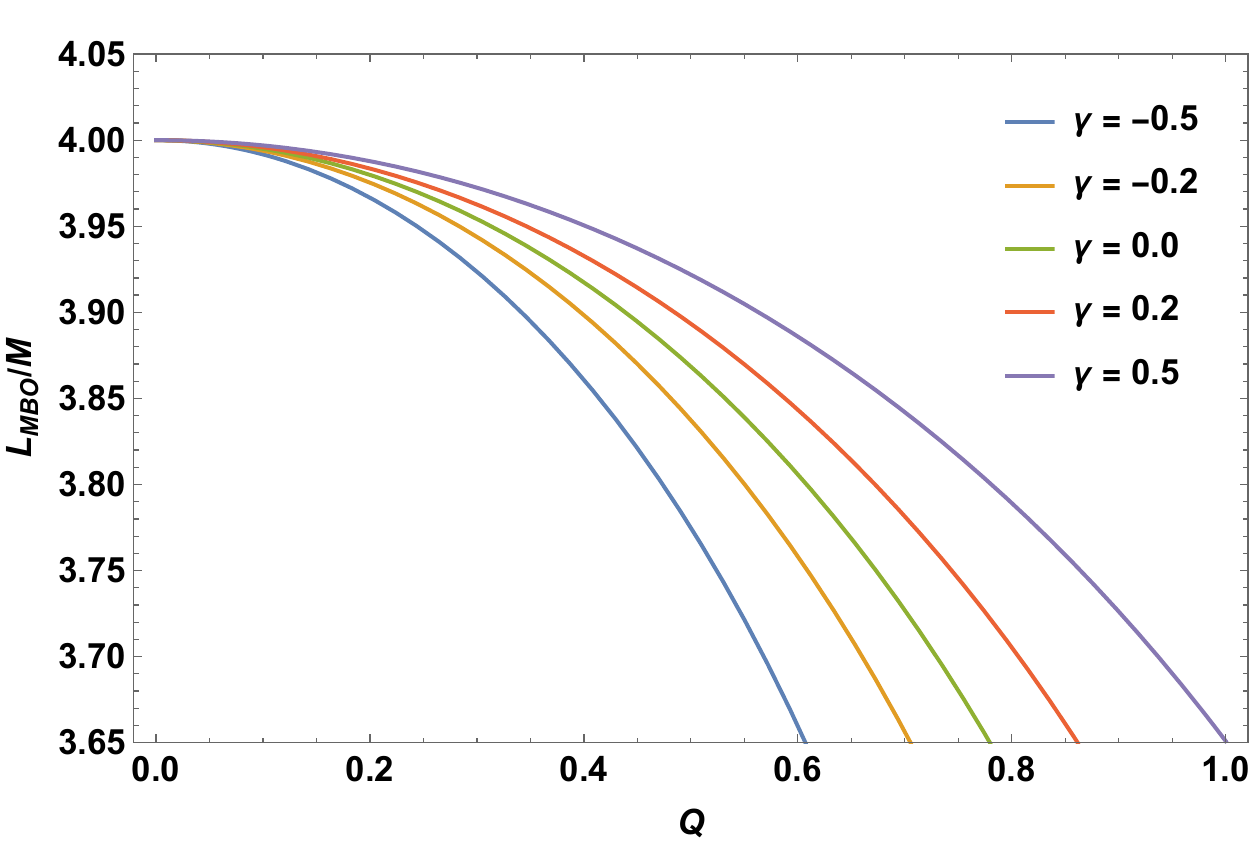}
\includegraphics[scale=0.4]{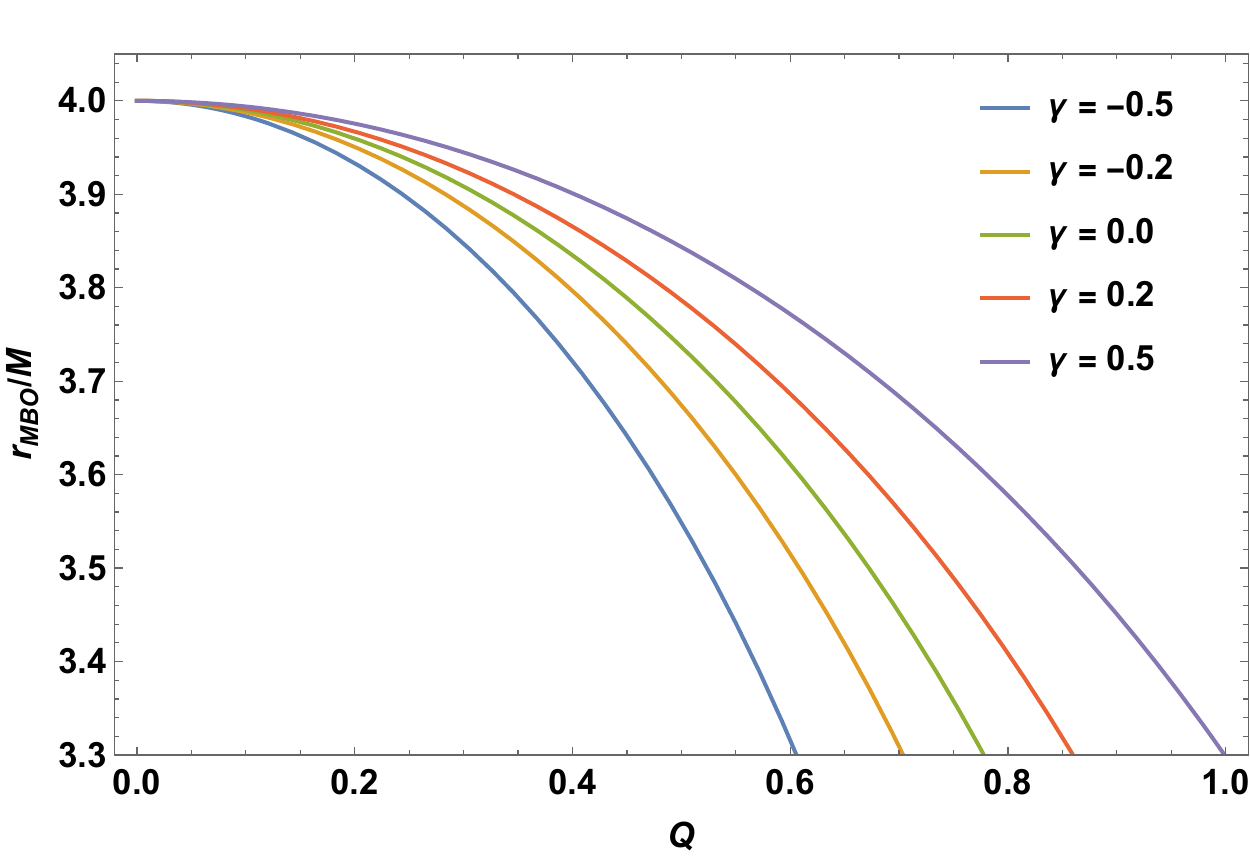}
\caption{The plot shows the dependence of the orbital angular momentum (left panel) and the radius of the MBO (right panel) on the BH charge for the different values of the $\gamma$ parameter.}
\label{fig:mbo}
\end{figure*}
\begin{figure*}[!htb]
\includegraphics[scale=0.4]{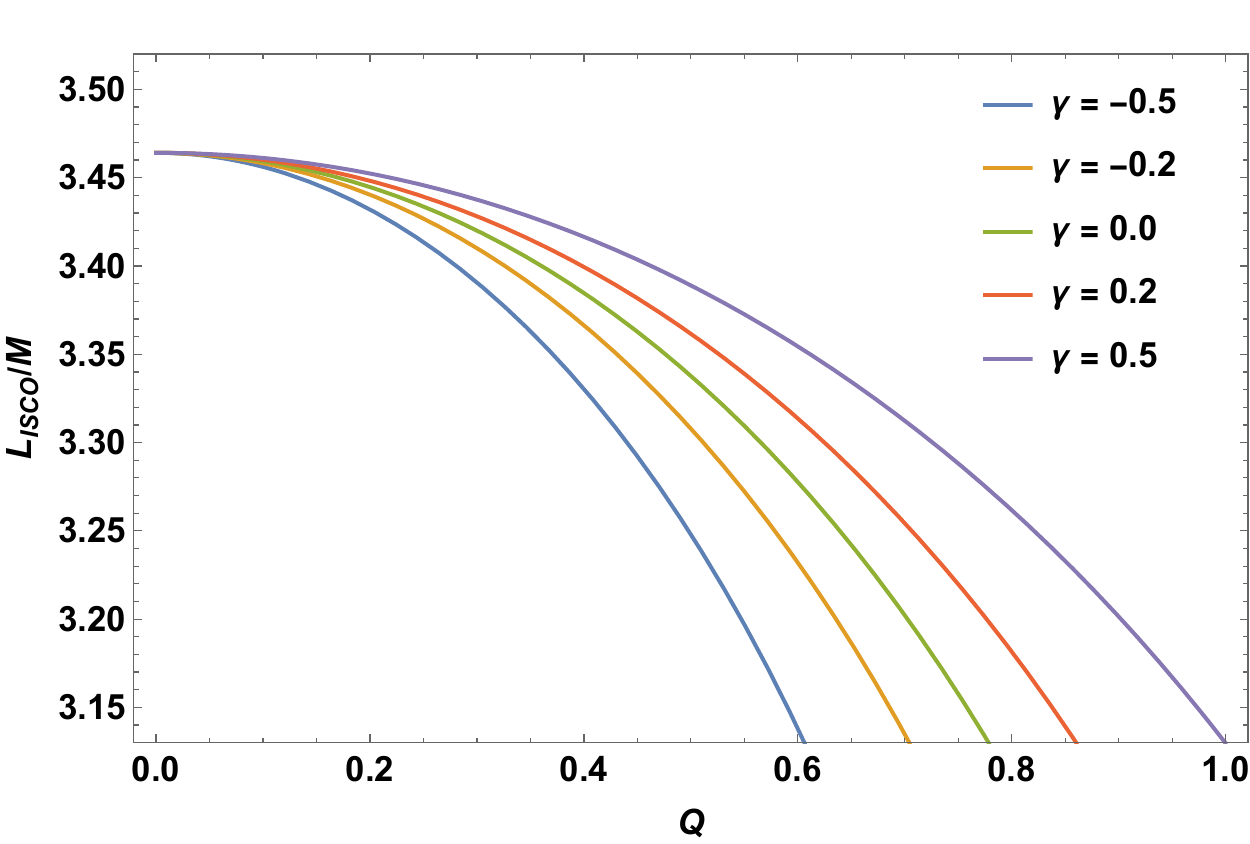}
\includegraphics[scale=0.4]{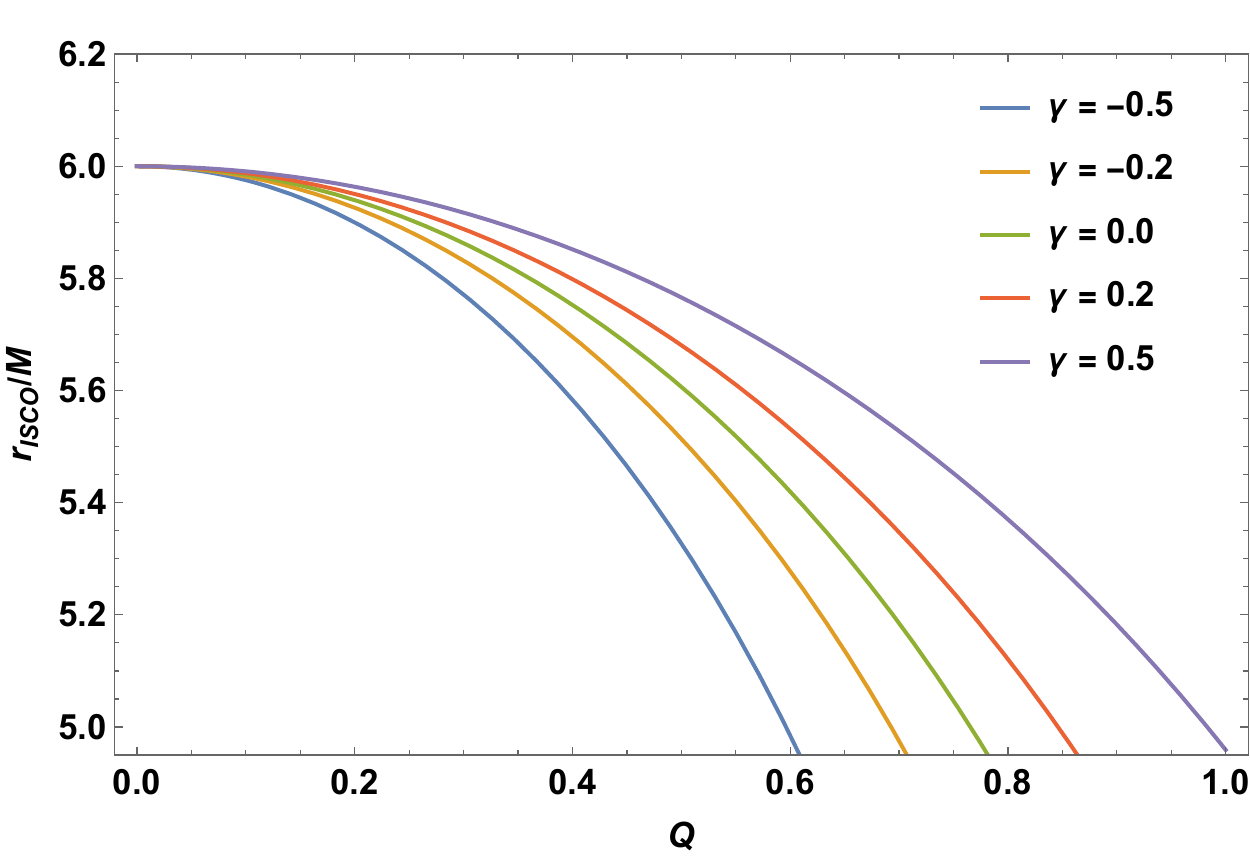}
\includegraphics[scale=0.4]{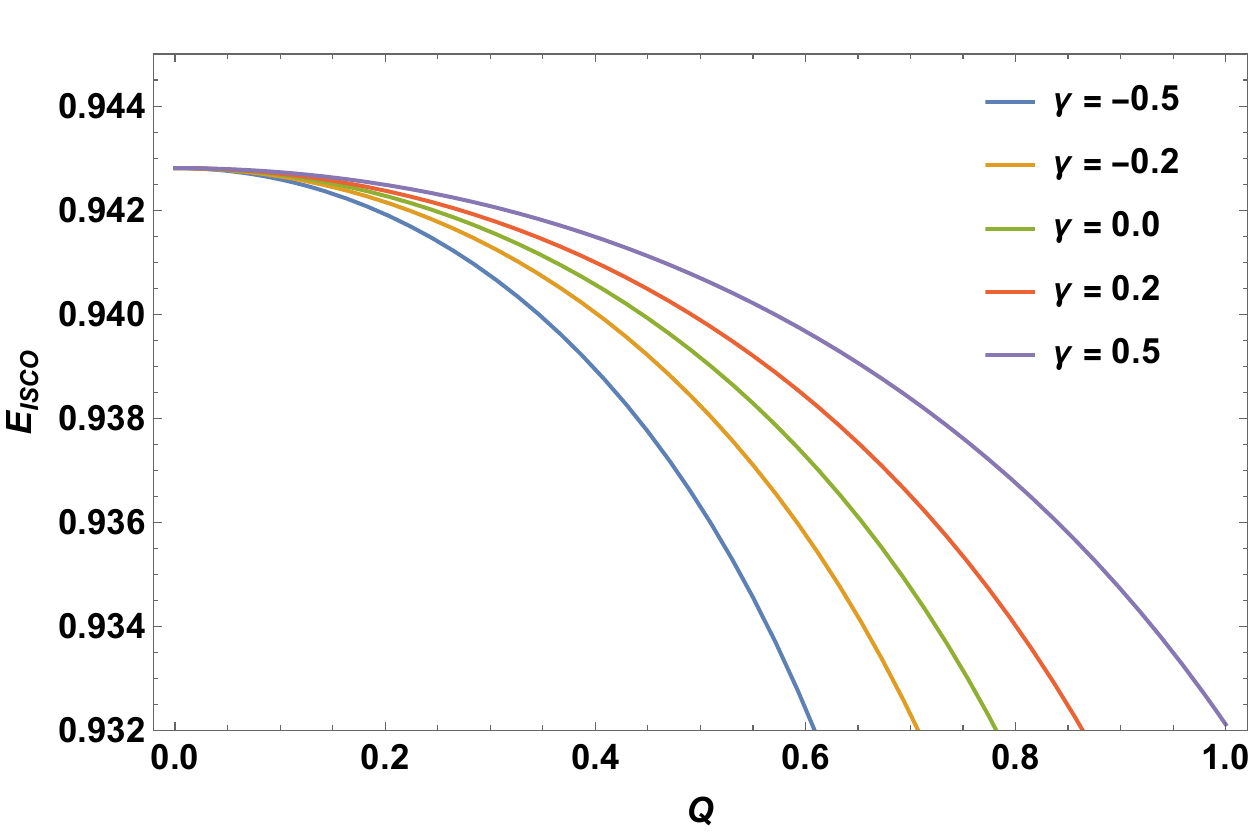}
\caption{The plot shows the dependence of the orbital angular momentum (top-left panel) and the radius of the ISCO (top-right panel) on the BH charge for the different values of the $\gamma$ parameter. Bottom panel illustrates the dependence of the ISCO energy on the BH charge $Q$ for different values of the $\gamma$ parameter. }
\label{fig:isco}
\end{figure*}
\begin{figure*}
\includegraphics[scale=0.5]{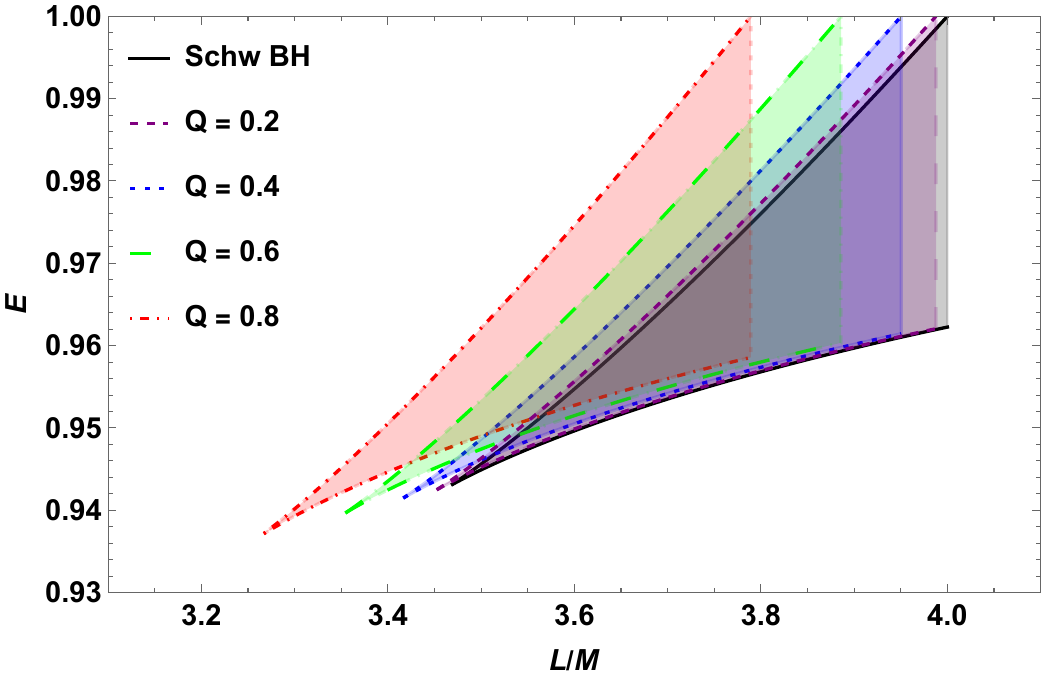}
\includegraphics[scale=0.5]{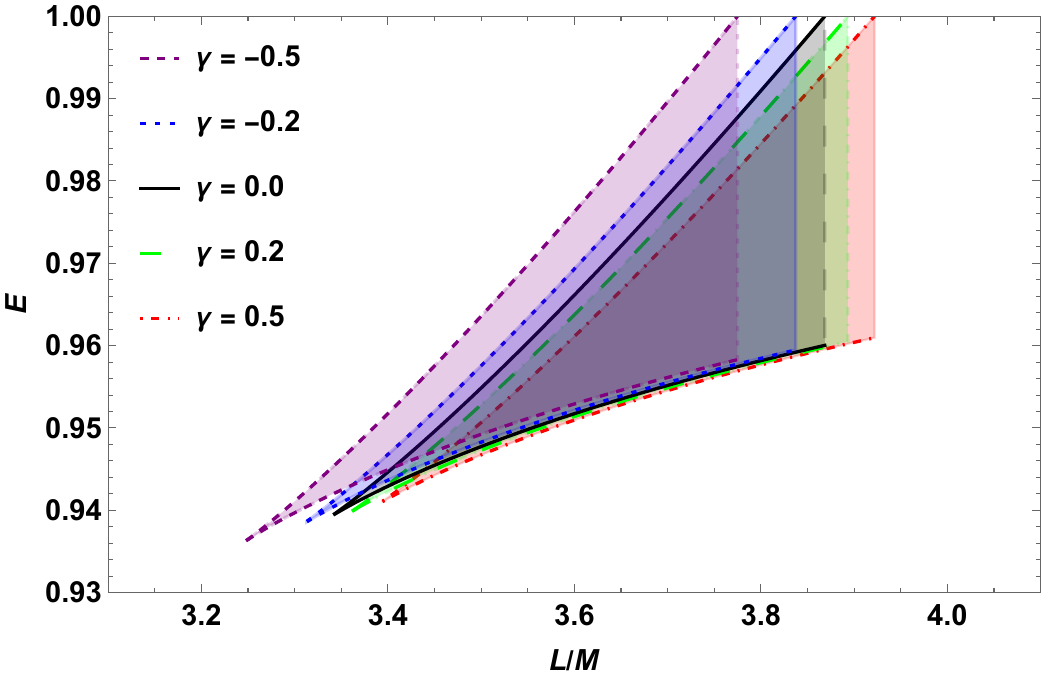}
\caption{The figure illustrates the allowed parameter space of the orbital angular momentum and energy for the bound orbits around the ModMax BH with different values of the BH charge and screening parameter. Here, we set $\gamma=0.5$ and $Q=0.5$ for the left and right panels, respectively.}
\label{fig:EL}
\end{figure*}
\begin{figure*}
    \includegraphics[scale=0.4]{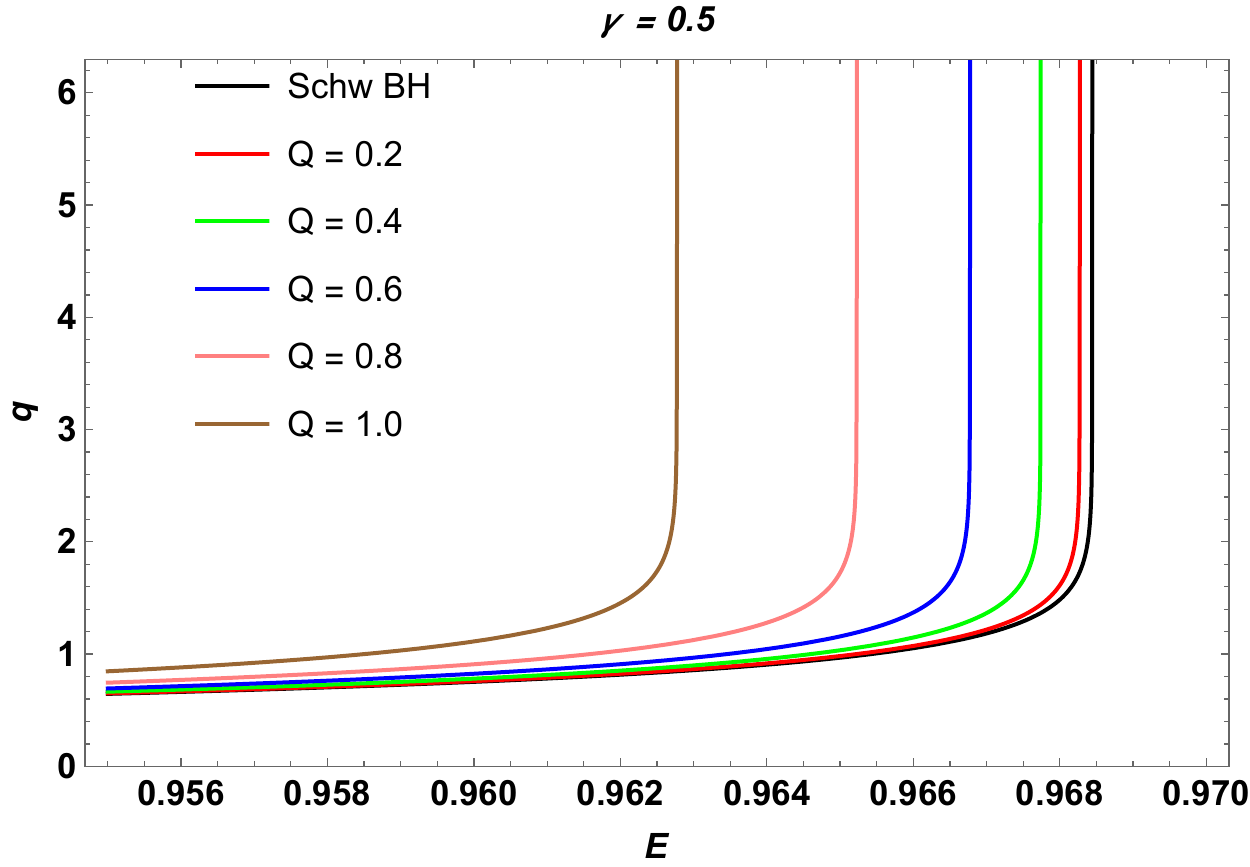}
    \includegraphics[scale=0.4]{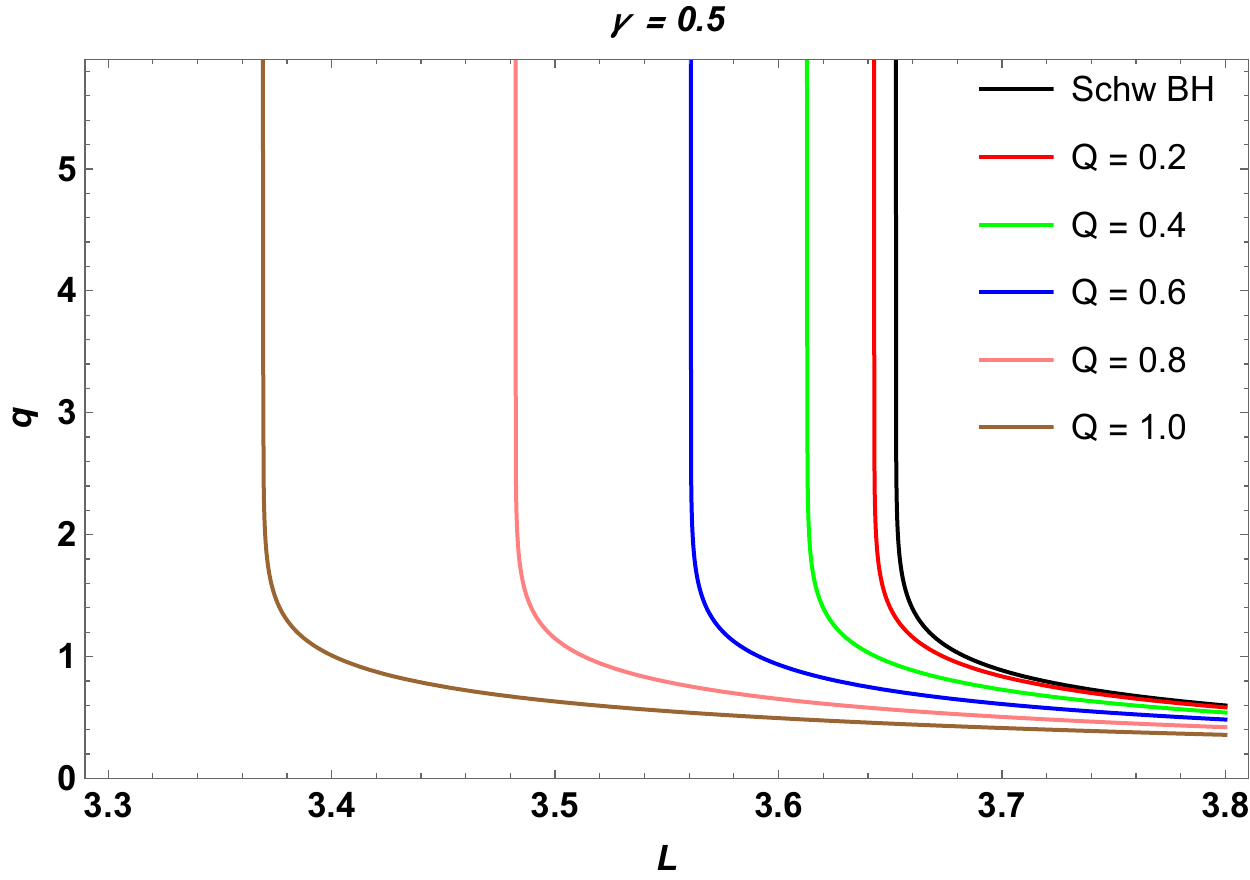}
    \caption{The figure demonstrates the dependence of the rational number $q$ on the energy (left panel) and orbital angular momentum (right panel) for different values of the BH charge. Here, we set $\gamma=0.5$ for both panels, $L=\frac{1}{2}(L_{MBO}+L_{ISCO})$ and $E=0.96$ for the left and right panels, respectively. }
    \label{q}
\end{figure*}
\begin{figure*}[htbp]
    \centering
    \includegraphics[width=0.32\textwidth]{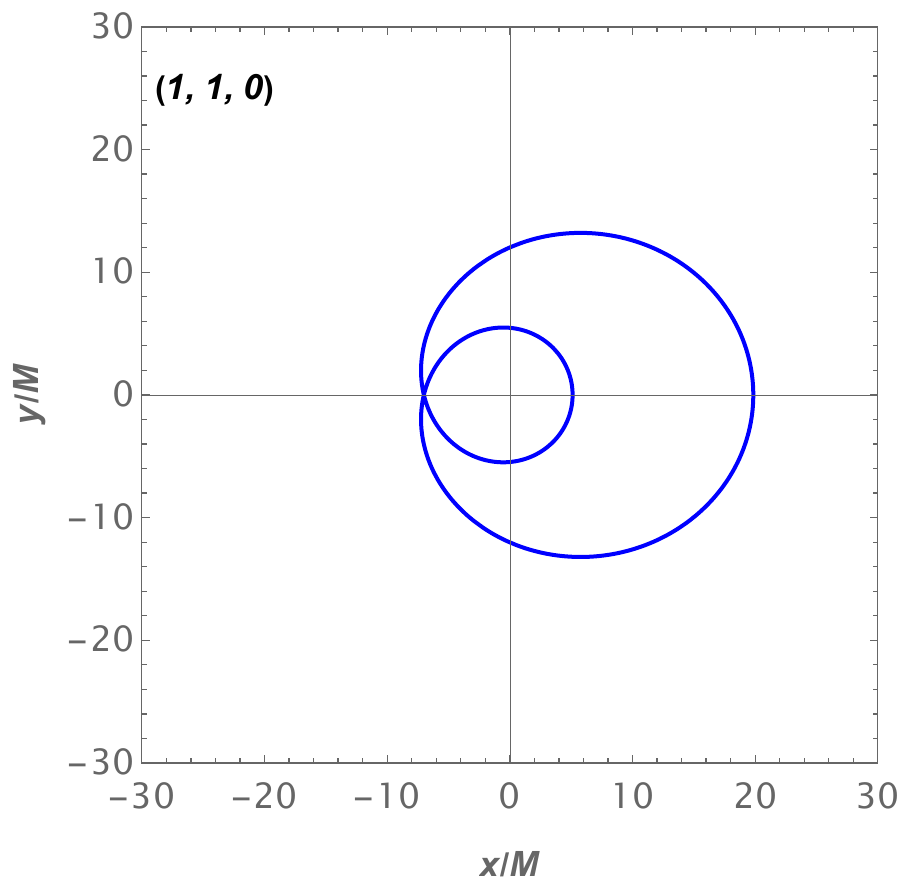} \hfill
    \includegraphics[width=0.32\textwidth]{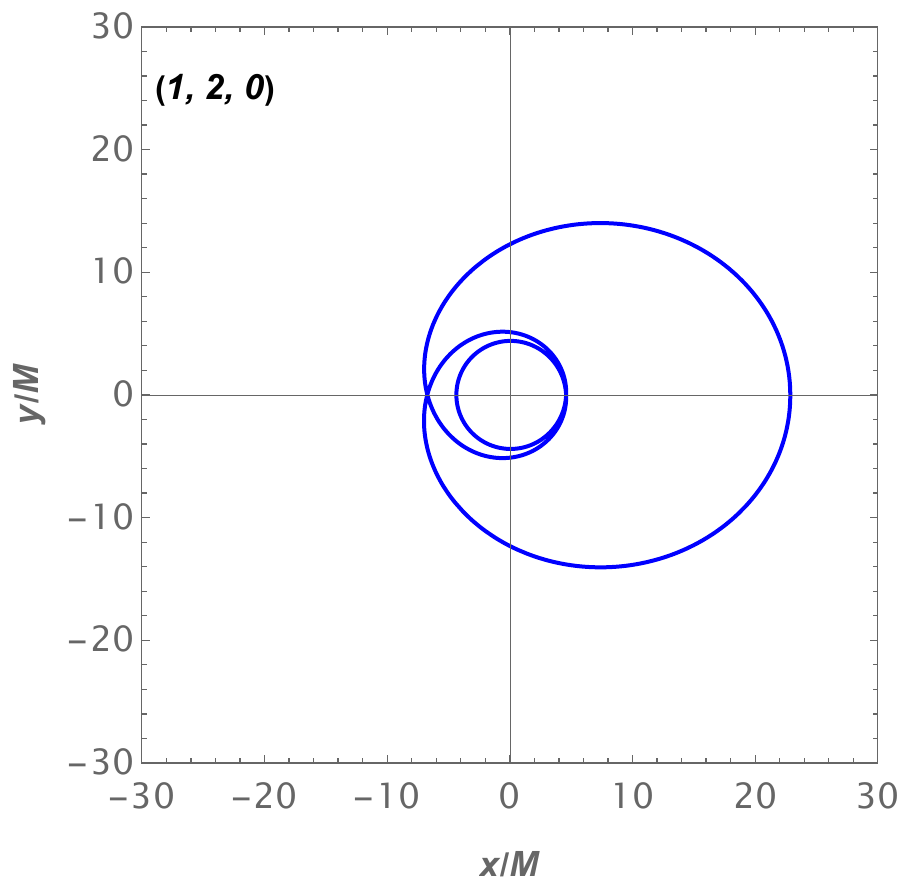} \hfill
    \includegraphics[width=0.32\textwidth]{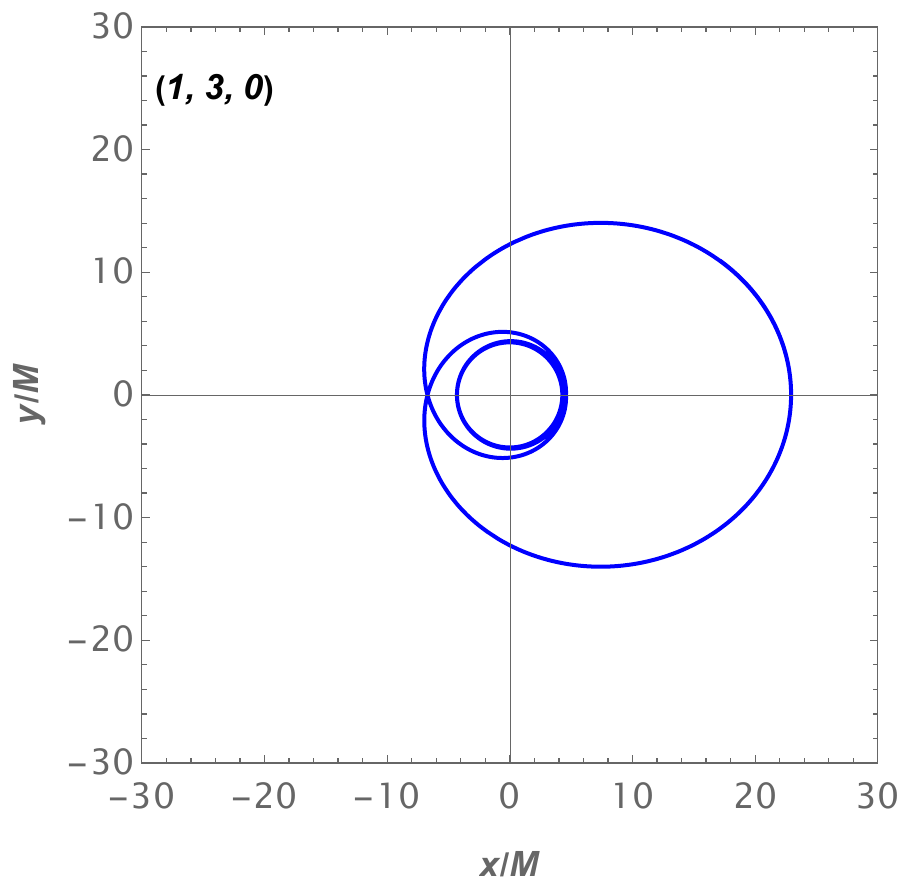} \\
    
    \vspace{0.2cm} 
    \includegraphics[width=0.32\textwidth]{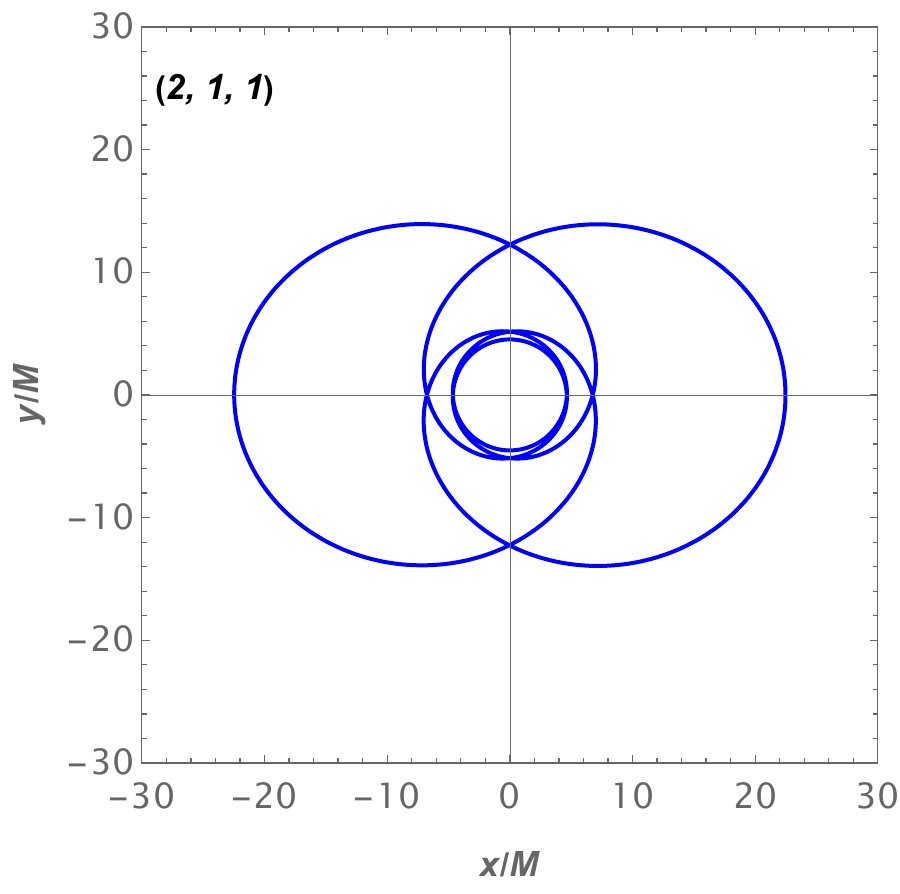} \hfill
    \includegraphics[width=0.32\textwidth]{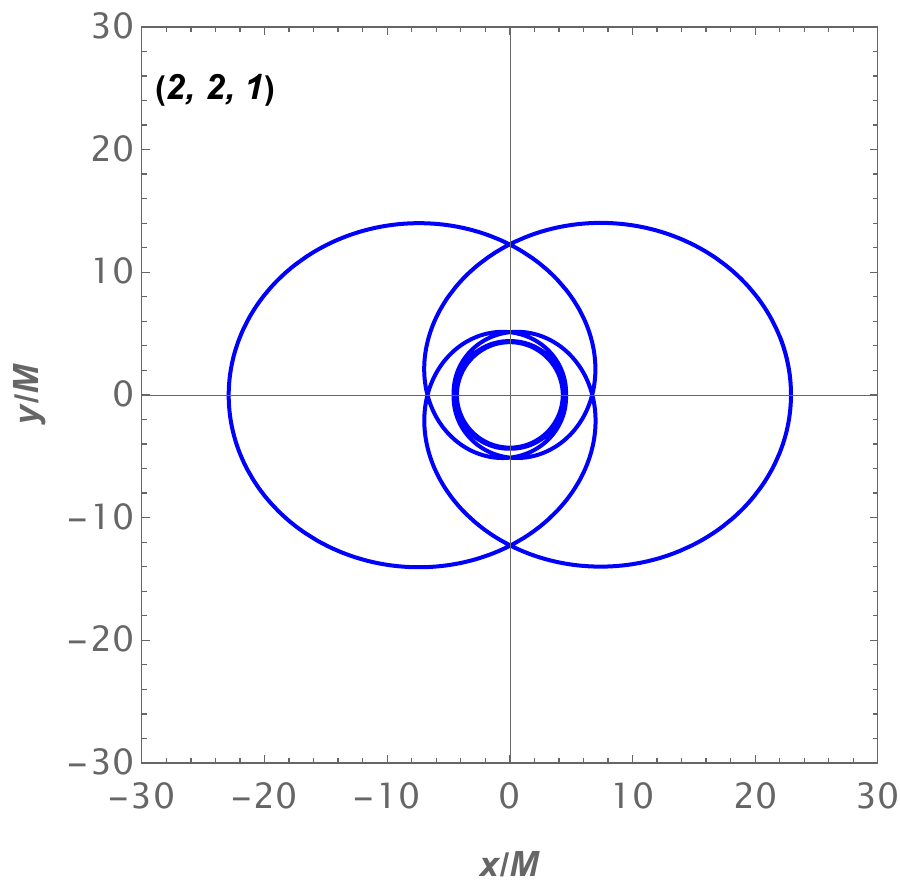} \hfill
    \includegraphics[width=0.32\textwidth]{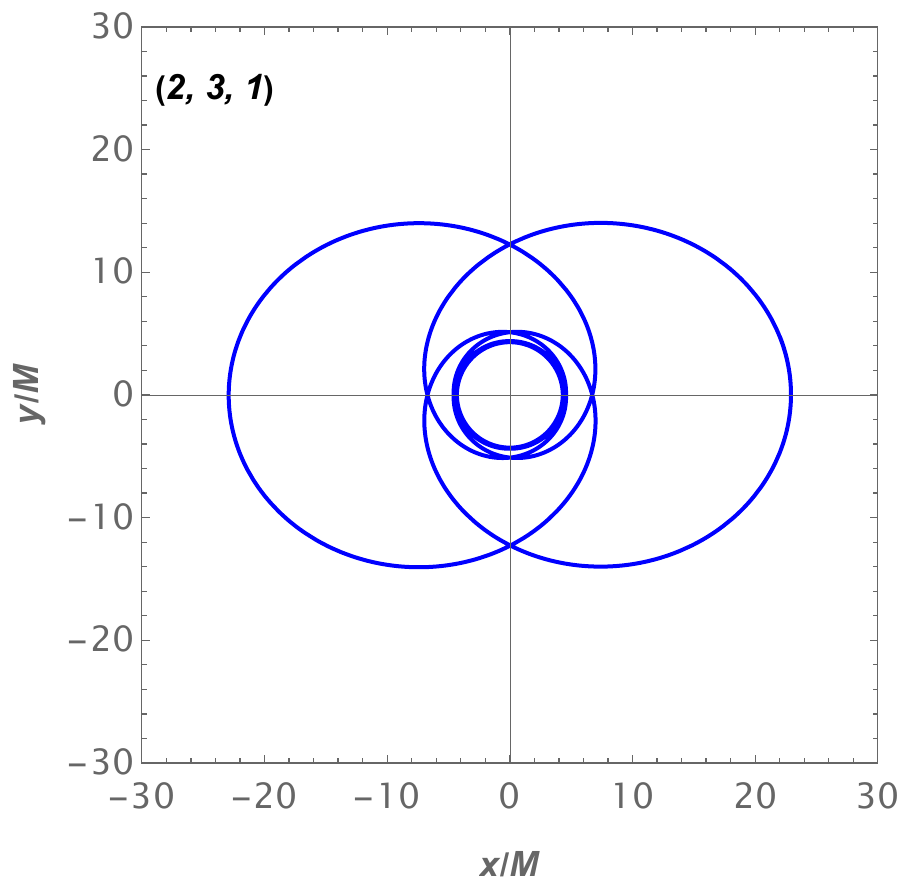} \\
    
    \vspace{0.2cm} 
    \includegraphics[width=0.32\textwidth]{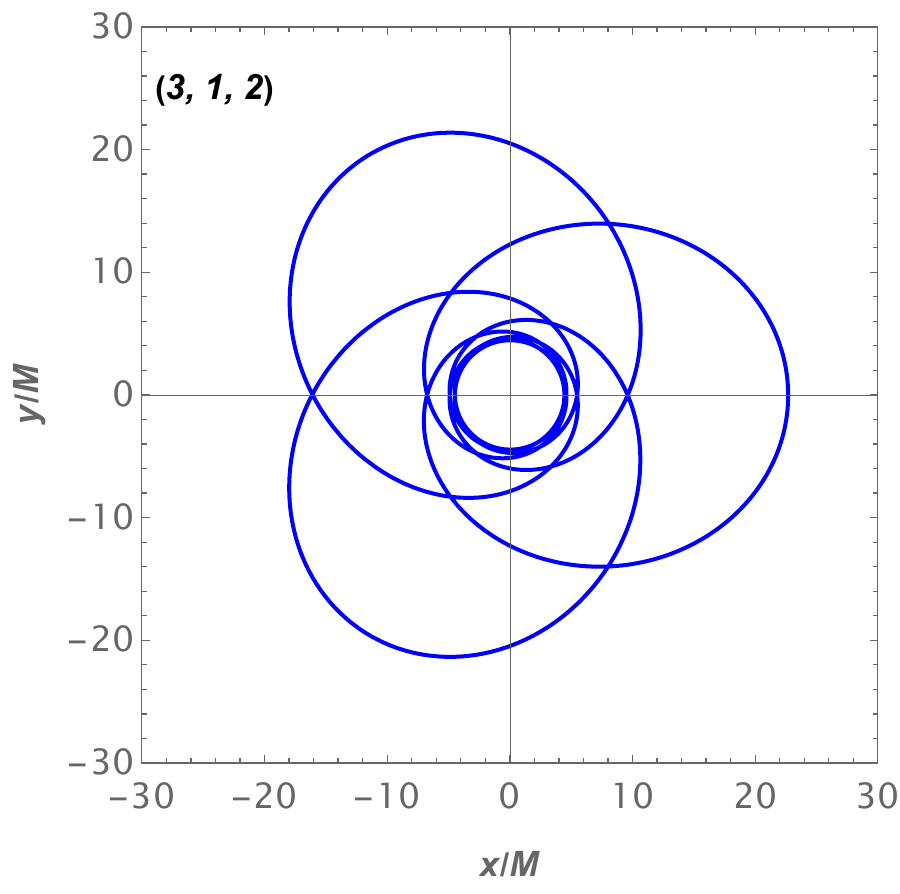} \hfill
    \includegraphics[width=0.32\textwidth]{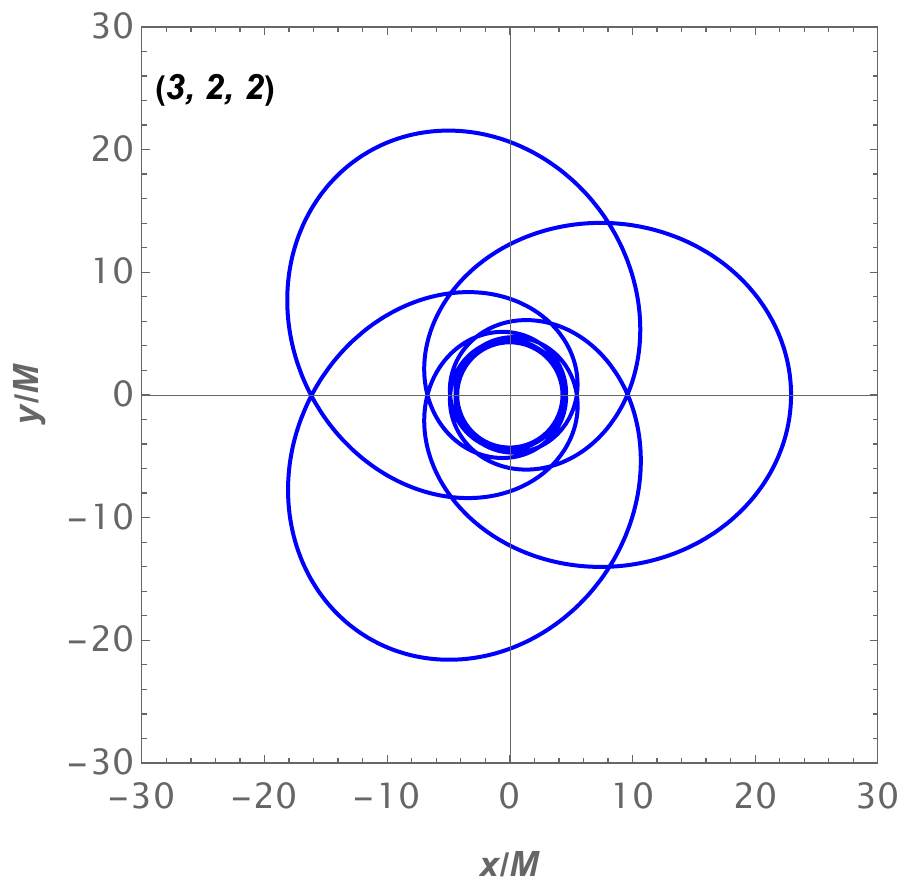} \hfill
    \includegraphics[width=0.32\textwidth]{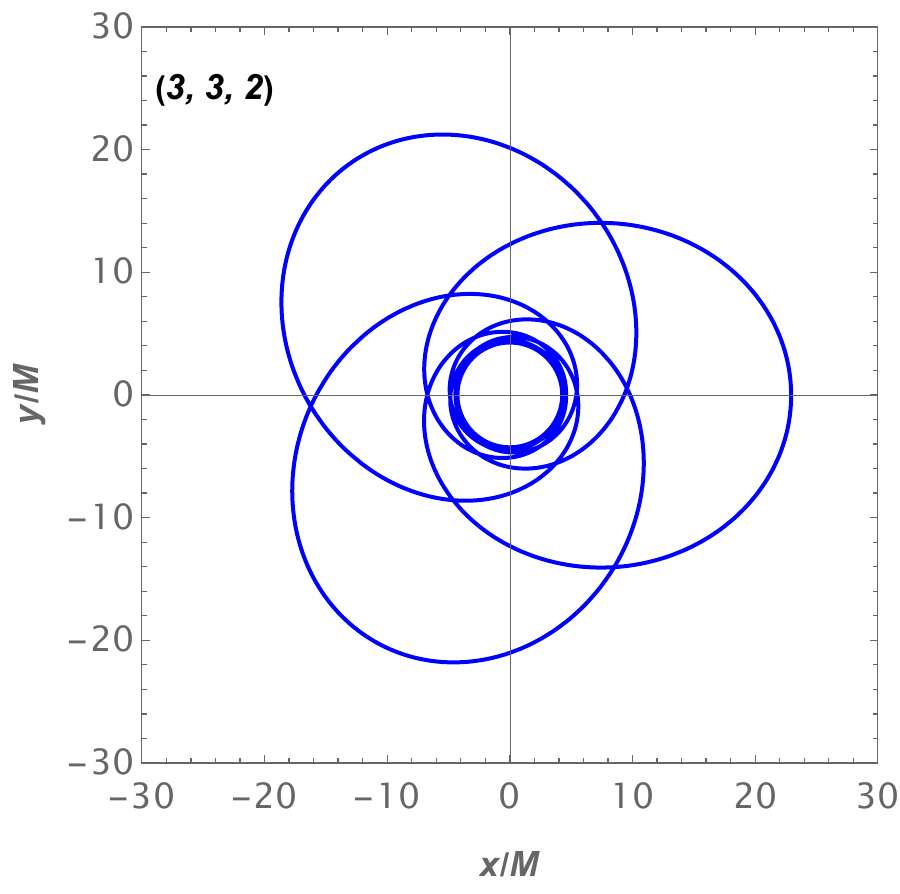} \\
    
    \vspace{0.2cm} 
    \includegraphics[width=0.32\textwidth]{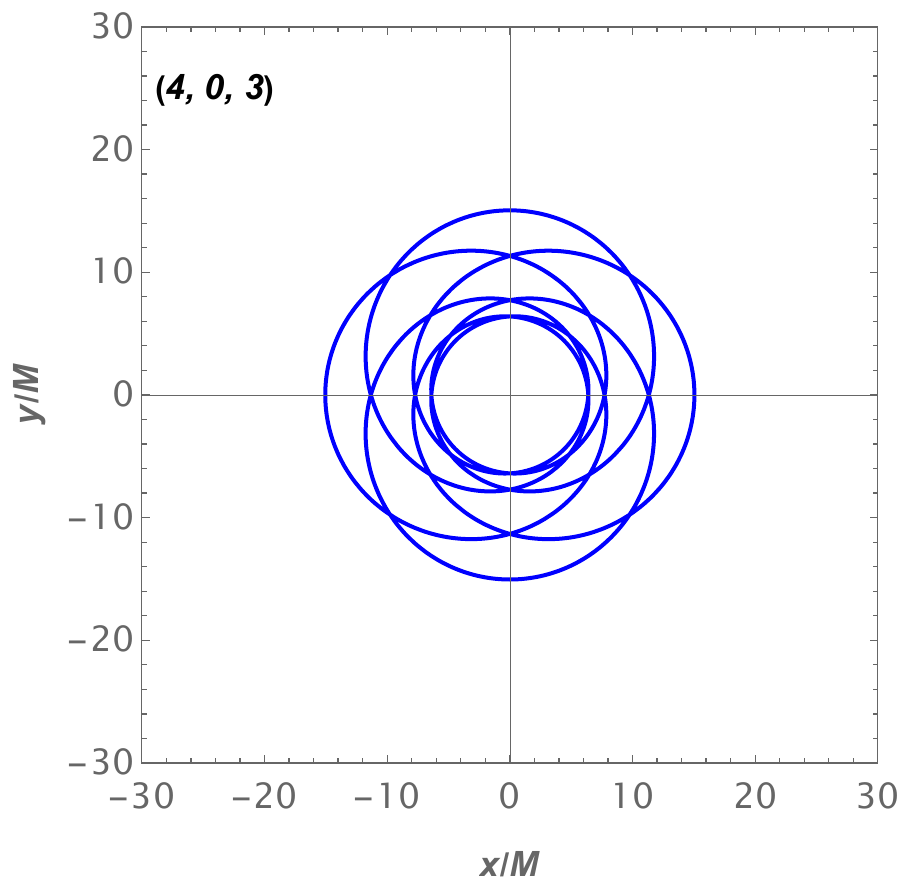} \hfill
    \includegraphics[width=0.32\textwidth]{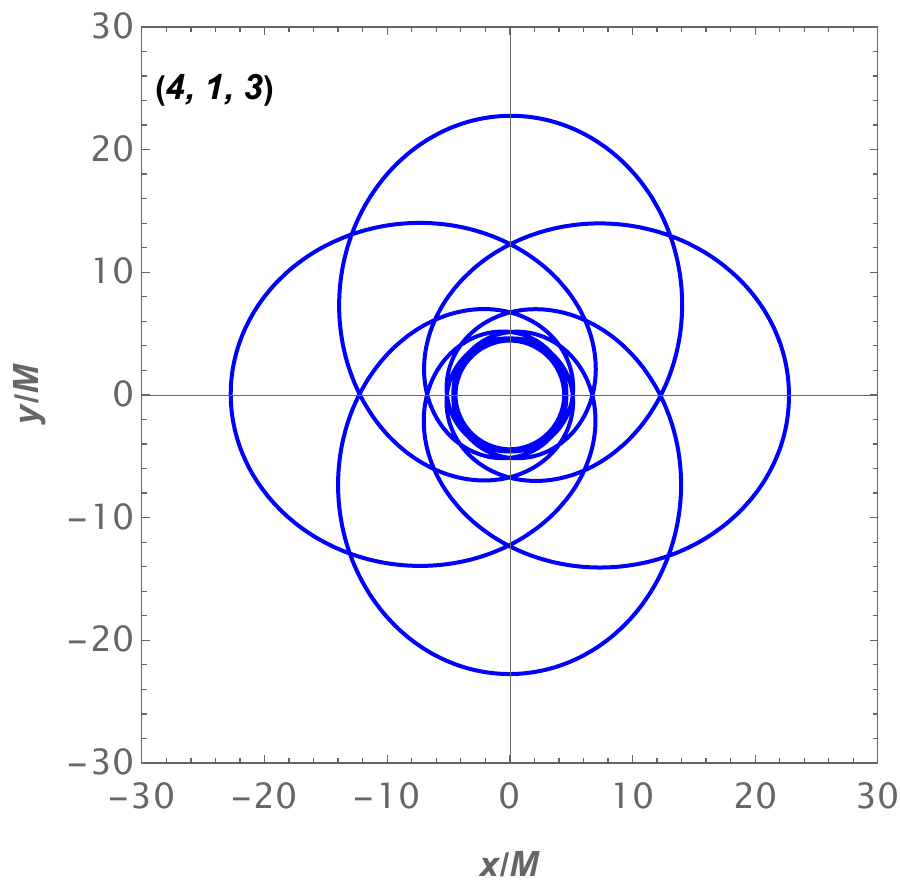} \hfill
    \includegraphics[width=0.32\textwidth]{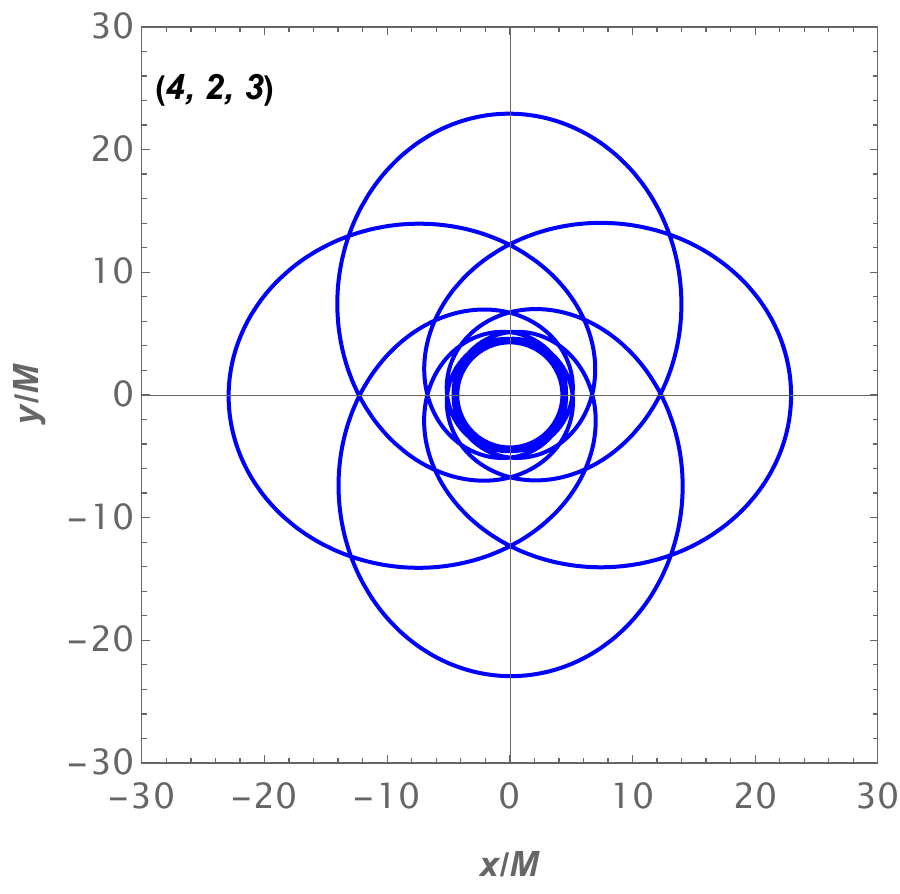}
    
    \caption{The periodic orbits for different combinations of integers $(z,w,v)$ around the ModMax BH. Here, we have set $Q=0.4$, $\gamma=0.5$ and $L=\frac{1}{2}(L_{MBO}+L_{ISCO})$.}
    \label{fig:periodicL}
\end{figure*}
\begin{figure*}[htbp]
\begin{subfigure}[b]{0.45\textwidth}
\includegraphics[width=\textwidth]{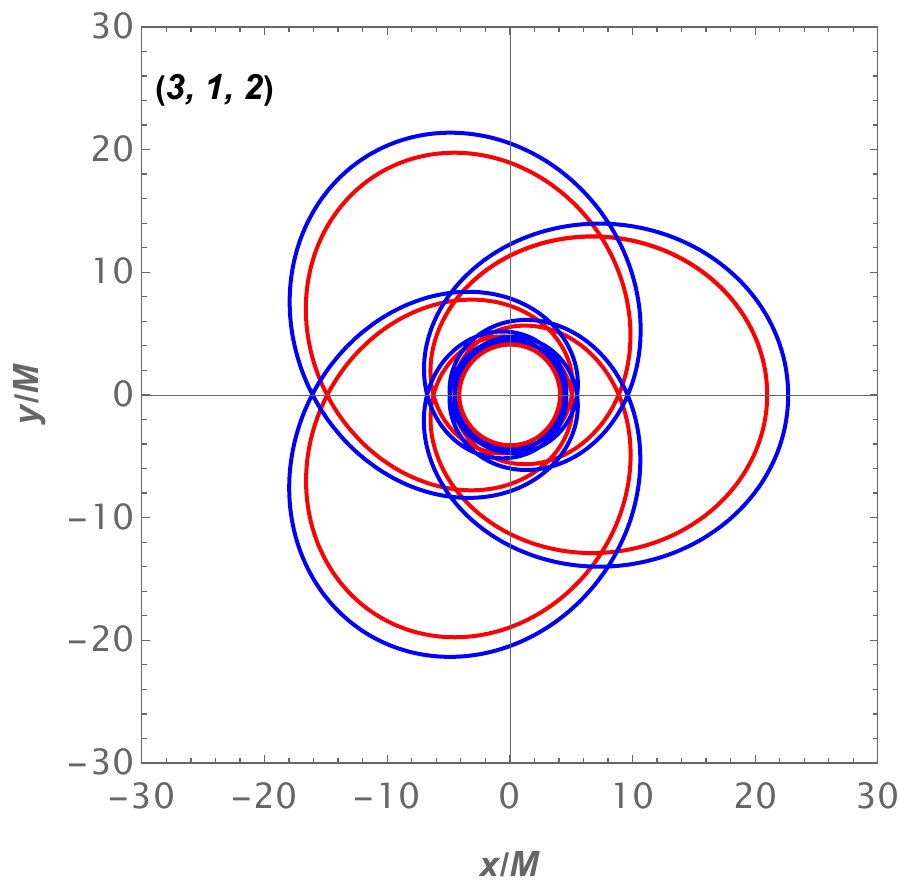}
\end{subfigure}
\begin{subfigure}[b]{0.52\textwidth}
\includegraphics[width=\textwidth]{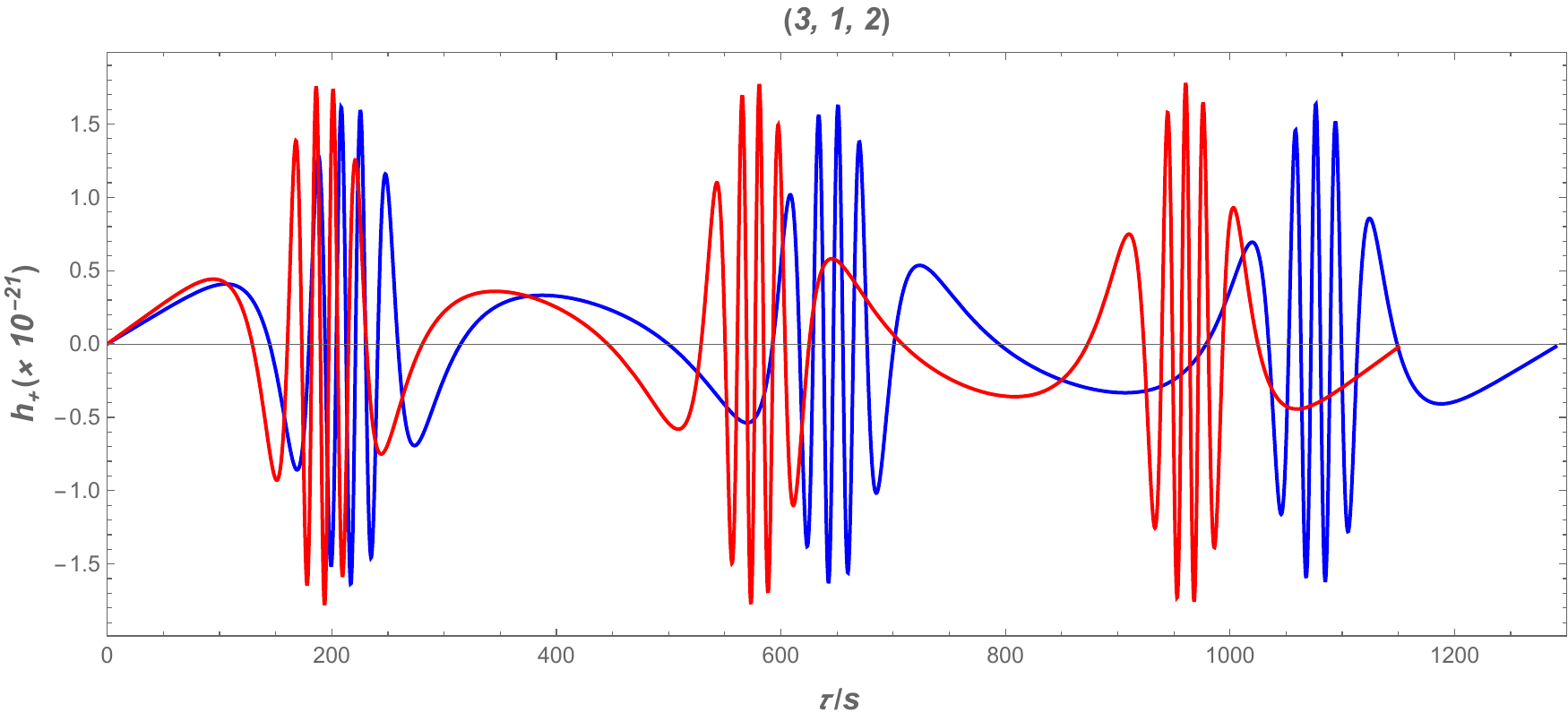}
\includegraphics[width=\textwidth]{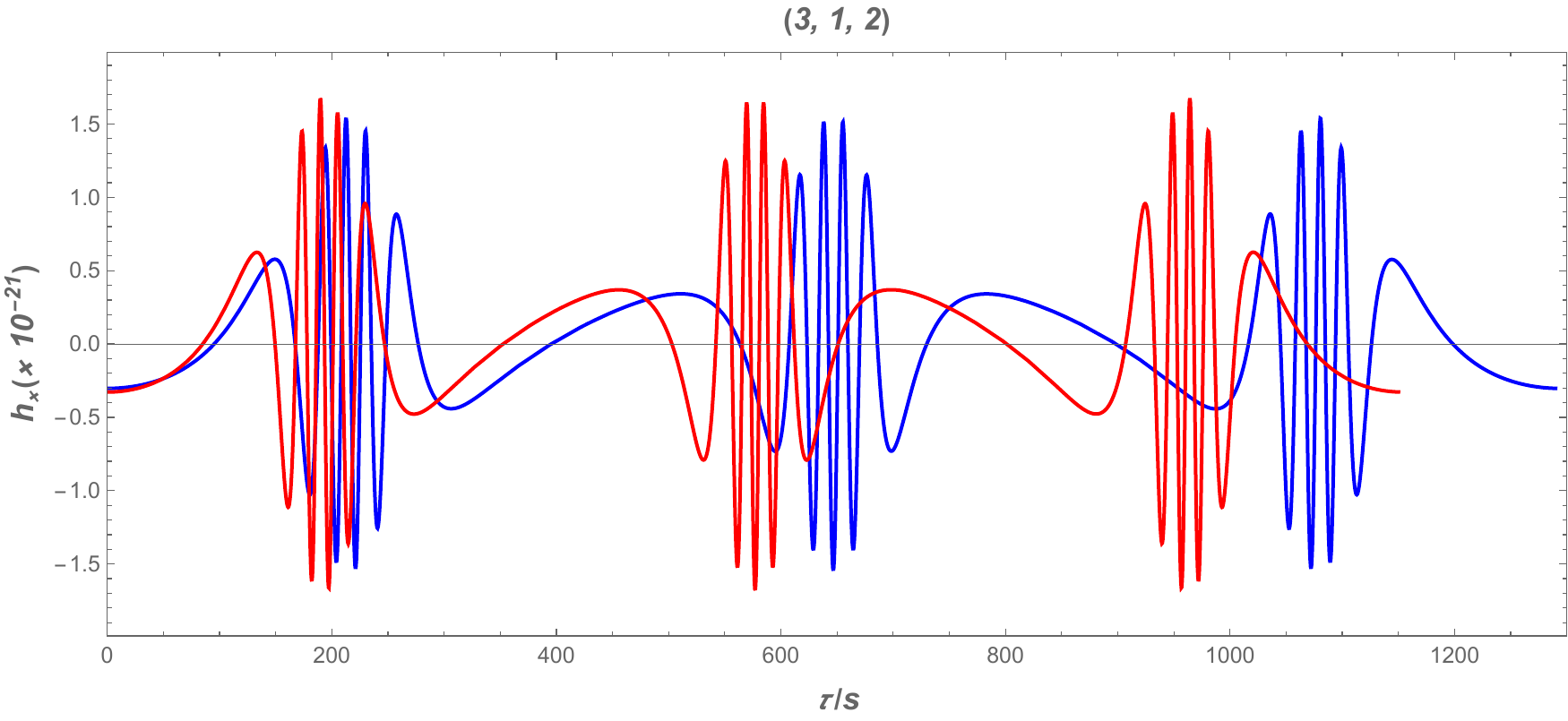}    
\end{subfigure} 
\caption{The plot demonstrates the $(3,1,2)$ periodic orbit and its associated GW signal of the EMRI system, which consists of the supermassive ModMax BH and the small object around this BH. We set the mass of the SMBH and the small object as $M\sim 10^7 M_{\odot}$ and $m\sim 10M_{\odot}$, respectively. The red line represents $Q=0.8$, while the blue one refers to $Q=0.4$. Here, the screening factor is equal to $\gamma=0.5$.}
\label{fig:combination}
\end{figure*}
\begin{figure*}[htbp]
\begin{subfigure}[b]{0.45\textwidth}
\includegraphics[width=\textwidth]{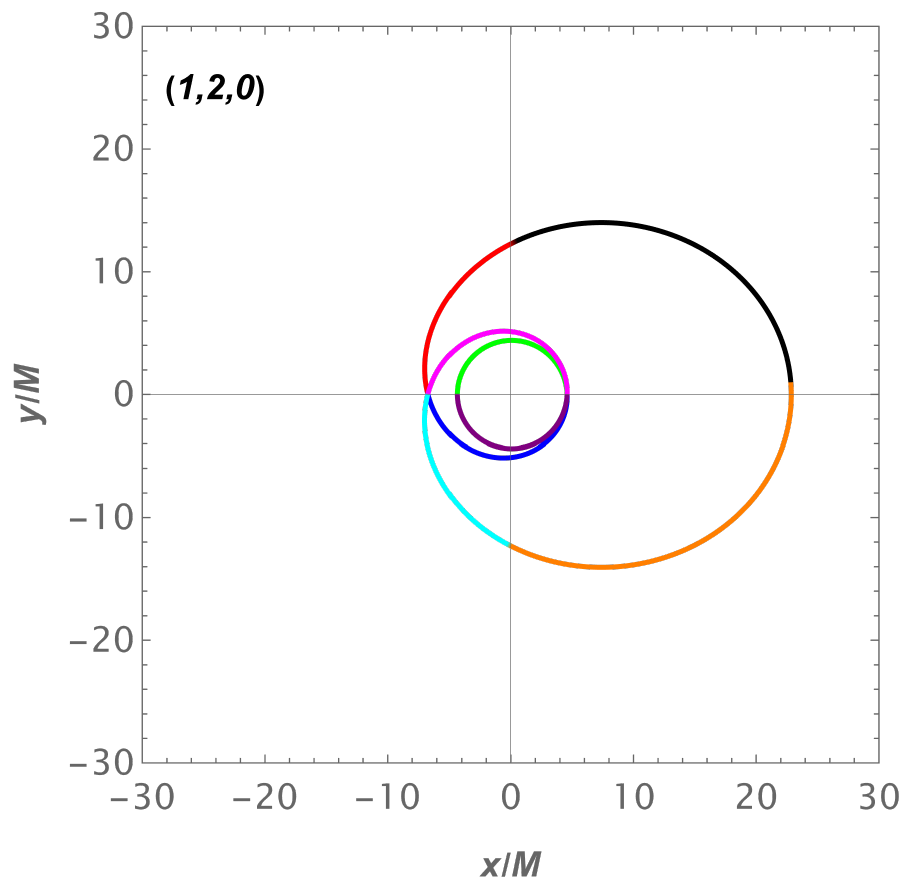}
\end{subfigure}
\begin{subfigure}[b]{0.52\textwidth}
\includegraphics[width=\textwidth]{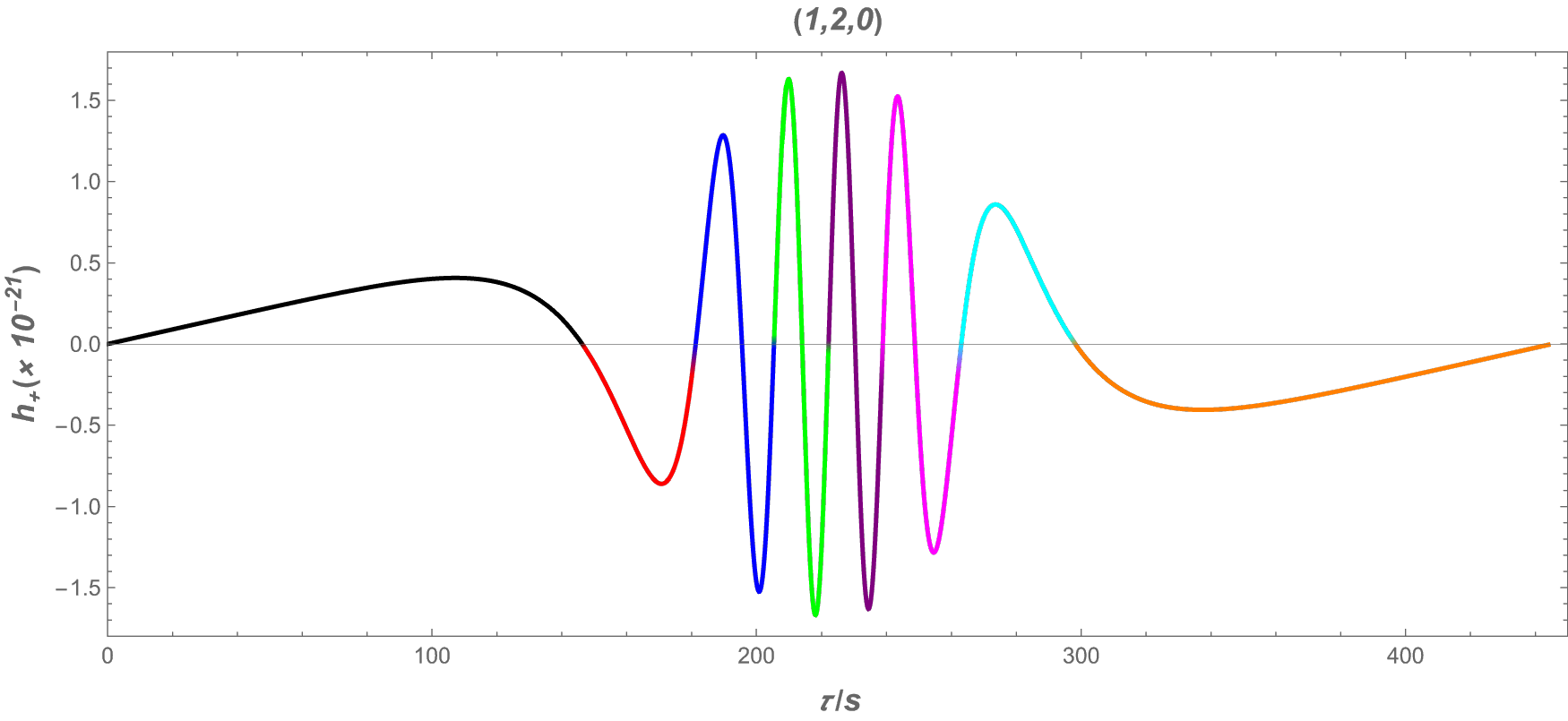}
\includegraphics[width=\textwidth]{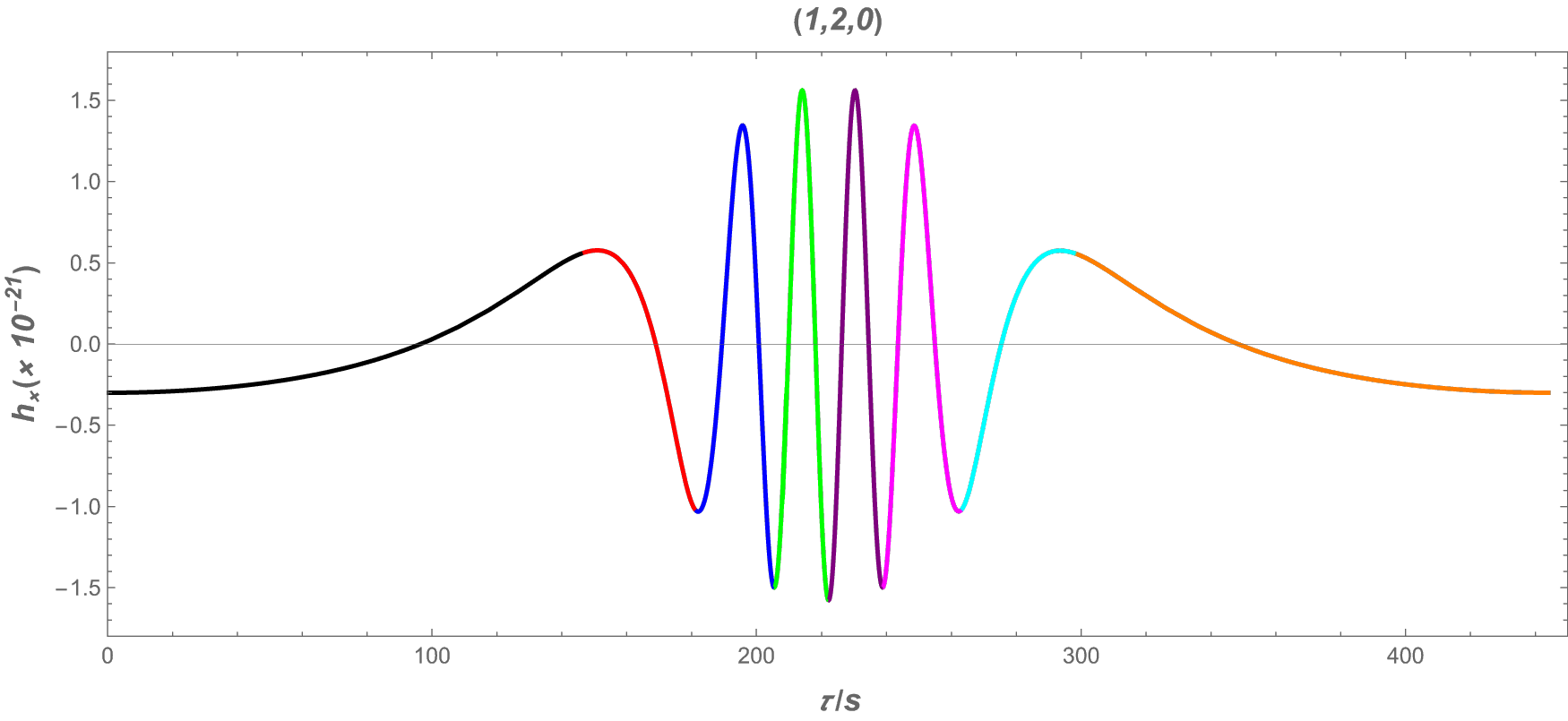}    
\end{subfigure} 
\caption{The plot demonstrates the $(1,2,0)$ periodic orbit and its associated GW signal of the EMRI system, which consists of the supermassive ModMax BH and the small object around this BH. We set the mass of the SMBH and the small object as $M\sim 10^7 M_{\odot}$ and $m\sim 10M_{\odot}$, respectively. Here, the screening factor is equal to $\gamma=0.5$. Also, each segment is highlighted through different colors.} 
\label{fig:combination2}
\end{figure*}
\begin{figure*}[htbp]
\includegraphics[width=0.85\textwidth]{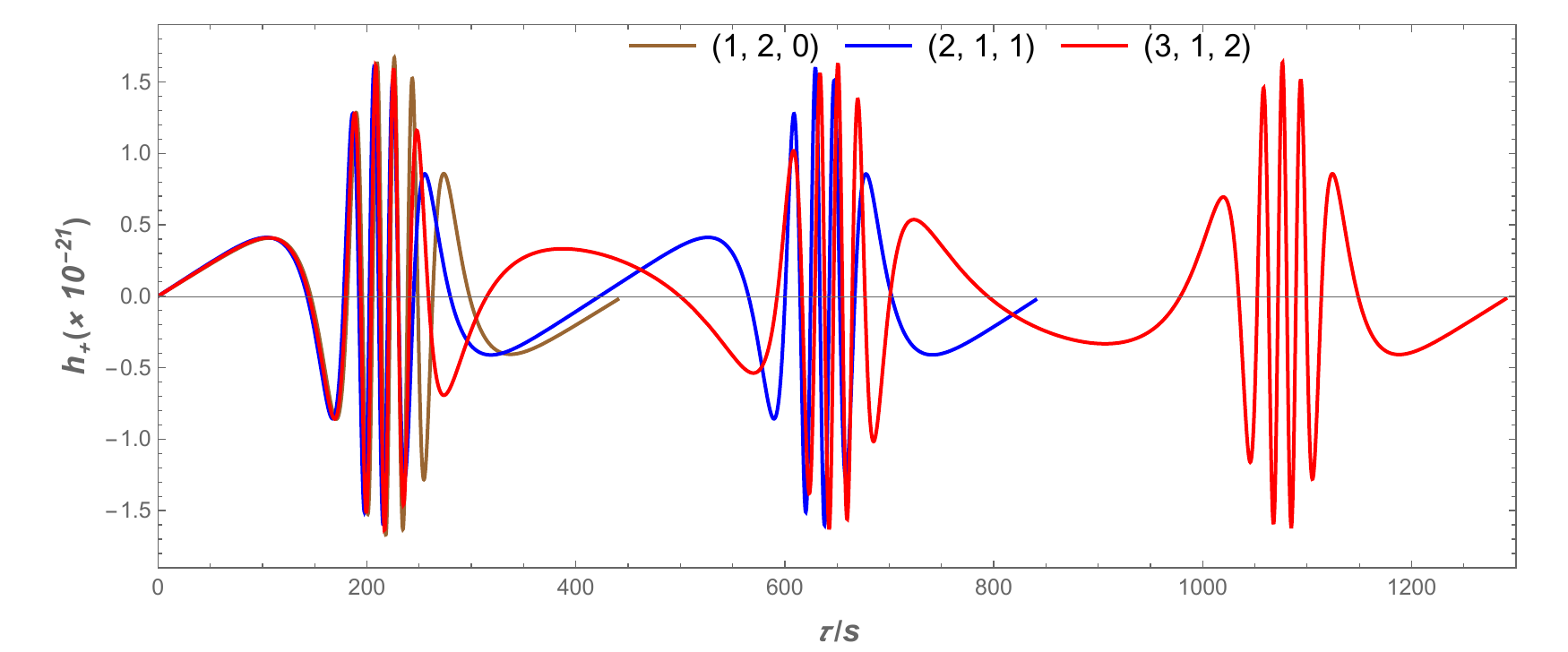}
\includegraphics[width=0.85\textwidth]{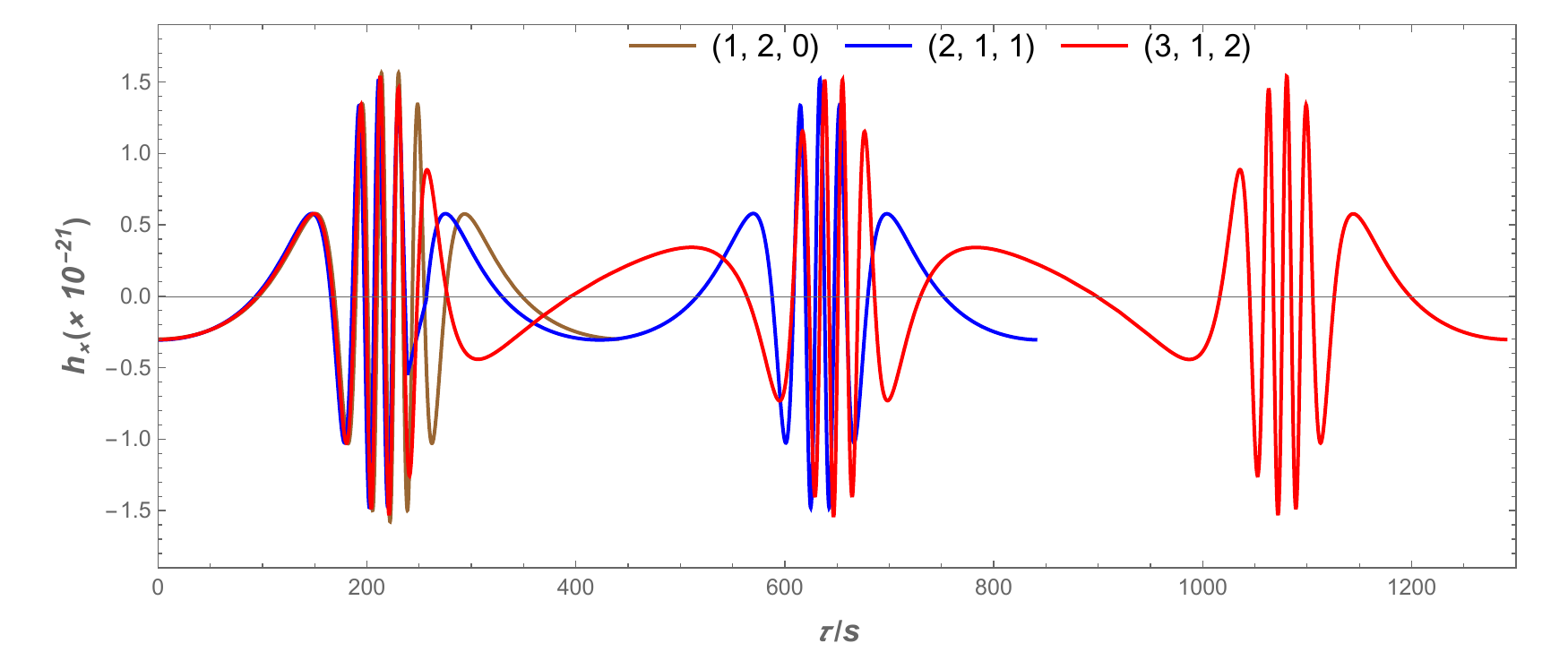}
\caption{The graph illustrates the distinct gravitational waveforms generated by various periodic orbits. The top panel presents the $h_{+}$, whereas the bottom panel depicts $h_{\times}$. The mass of the SMBH and the small object are set to $M\sim 10^7 M_{\odot}$ and $m\sim 10M_{\odot}$, respectively. Here, we set $Q=0.4$ and $\gamma=0.5$.}
\label{fig:GW}
\end{figure*}
In this section, we briefly review the spacetime of the dyonic ModMax BH and geodesic orbits. The line element of this spacetime can be written in the following form~\cite{SIAHAAN2025139479}
\begin{equation}
ds^2=-f(r)dt^2+\frac{1}{f(r)}dr^2+r^2(d\theta^2+\sin^2{\theta} d\phi^2)\ ,
\end{equation}
with
\begin{eqnarray}
    f(r)=1-\frac{2M}{r}+\frac{Q^2}{r^2}e^{-\gamma}\, , 
\end{eqnarray}
with $Q^2=Q_e^2+Q_m^2$, where $Q_e$ and $Q_m$ respectively refer to the electric and magnetic charges of the BH, while $\gamma$ to the screening factor. In this work, we use $Q$ without separating it into two parts. It should be noted that when $\gamma=0$, we can recover the RN BH. If $Q=\gamma=0$, the Schwarzschild spacetime can be obtained. We plot the radial dependence of the metric function $f(r)$ for different values of the BH charge and the screening parameter $\gamma$ in Fig.~\ref{fig:f(r)}. Using $f(r)=0$, one can find the event horizon of the ModMax BH, and we can see from Fig.~\ref{fig:f(r)} that the structure of the event horizon changes under the influence of the spacetime parameters. To be more informative, we plot the phase diagram in Fig.~\ref{fig:phasediagram}, which shows the existence of the dyonic ModMax BH in the $(Q,r)$ plane for different values of the parameter $\gamma$. The blue region with different opacities corresponds to the slice of the parameter space where $f(r,Q)\leq 0$, indicating the presence of dyonic ModMax BH. 

Moreover, here, we investigate the test particle dynamics around the dyonic ModMax BH. We can write the Lagrangian of the test particles as~\cite{1983mtbh.book.....C}
\begin{equation}
\mathcal{L}=\frac{1}{2}m\,g_{\mu\nu}\,\frac{dx^{\mu}}{d\tau}\,\frac{dx^{\nu}}{d\tau},
\end{equation}
where $\tau$ and $m$ refer to the proper time and mass of the test particle, respectively. Using the above equation, we can write the generalized momentum of the 
test particle by setting $m=1$ for simplicity as
\begin{equation}
    p_{\mu}=\frac{\partial {\cal L}}{\partial \dot{x}^{\mu}}=g_{\mu \nu}\dot{x}^{\nu}\, .
\end{equation}
Then, the equations of motion of the test particles can be obtained by using the above equation as follows~\cite{1983mtbh.book.....C}
\begin{eqnarray}\label{eq:eqmotion}
p_{t} &=& -\left(1-\frac{2M}{r}+\frac{Q^2}{r^2}e^{-\gamma}\right)\dot{t}=-E \, , \nonumber\\
p_{\phi} &=& r^{2}\sin^{2}\theta\dot{\phi}=L \, , \nonumber \\
p_{r} &=& \left(1-\frac{2M}{r}+\frac{Q^2}{r^2}e^{-\gamma}\right)^{-1}\dot{r} \, , \nonumber\\
p_{\theta} &=& r^{2}\dot{\theta} \, ,
\end{eqnarray}
where $L$ and $E$ are the orbital angular momentum and the energy of the particle, respectively. In this work, the motion in the equatorial plane is considered ($\theta=\pi/2$). Following the normalization condition, $g_{\mu\nu}\dot{x}^{\mu}\dot{x}^{\nu}=-1$,
%
we write the radial equation of motion as 
\begin{eqnarray}
\dot{r}^{2}+V_{\rm eff}=E^{2}\, ,
\end{eqnarray}
with
\begin{equation}
V_{\rm eff}=\left(1-\frac{2M}{r}+\frac{Q^2}{r^2}e^{-\gamma}\right)\left(1+\frac{L^{2}}{r^{2}}\right)\, ,
\end{equation}
where $V_{eff}$ is the effective potential for the motion of the test particle. In Fig.~\ref{fig:eff}, we show the radial dependence of the effective potential for different values of the spacetime parameters and the orbital angular momentum. It can be seen from this figure that the values of the effective potential increase with the increase in the BH charge. In contrast, the values of the effective potential decrease with increasing $\gamma$ parameter. Moreover, there is an increase under the influence of the orbital angular momentum of the test particles. The core objective of our work is to investigate the periodic orbits around the dyonic ModMax BH. As a special class of bound orbits, periodic orbits require the particle's energy and angular momentum to fulfill specific conditions, which are given below \cite{Dadhich22a}:  
\begin{equation}
    L_{ISCO}\leq L \quad\mbox{and}\quad E_{ISCO}\leq E \leq E_{MBO}=1\, ,
\end{equation}
where $L_{ISCO}$ and $E_{ISCO}$ are the orbital angular momentum and the energy of the particle that is moving on ISCO, respectively. We need to explore MBO and ISCO to analyze the properties of the periodic orbits. The corresponding conditions to determine the MBO can be written in the following form
\begin{equation}
    V_{eff}=1\quad\mbox{and}\quad \frac{d V_{eff}}{dr}=0.
\end{equation}
Based on these conditions, we numerically compute the radius and orbital angular momentum of the MBO. In Fig.~\ref{fig:mbo}, we plot the dependence of the radius and orbital angular momentum on the BH charge for different values of the screening factor $\gamma$. It can be seen from this figure that the values of these quantities decrease with an increase in the BH charge. In contrast, there is an increase under the influence of the $\gamma$ parameter. We next investigate another significant bound orbit, the ISCO. The ISCO parameters can be determined under the following conditions
\begin{equation}
    \dot{r}=0, \quad \frac{d V_{eff}}{dr}=0 \quad\mbox{and}\quad \frac{d^2V_{eff}}{dr^2}=0.
\end{equation}
In Fig.~\ref{fig:isco}, we demonstrate the ISCO parameters, which are the radius, orbital angular momentum, and energy. One can see from this figure that the values of the ISCO parameters decrease as the BH charge increases. On the other hand, there is an increase under the influence of the $\gamma$ parameter. To provide more information, one can plot the space of allowed parameters for the energy and orbital angular momentum for bound orbits around the dyonic ModMax BH for different values of the BH charge and screening parameter $\gamma$ in Fig.~\ref{fig:EL}. One can see from the left panel of this figure that the region shifts to the left as the BH charge increases. In contrast, the region shifts to the right with an increase in the screening parameter $\gamma$. This means that when the energy is fixed, bound orbits around the dyonic ModMax BH with large values of charge have a lower boundary for the orbital angular momentum and vice versa for the screening parameter. 
\section{Periodic orbits around the dyonic ModMax black hole}
\label{periodic}
\renewcommand{\arraystretch}{1.2}
\begin{table*}[]
\centering
\resizebox{0.9\textwidth}{!}{
\begin{tabular}{c|c|c|c|c|c|c|c|c|c}
\hline
\hline
$Q$ & $L$ & $E_{(1,1,0)}$ & $E_{(1,2,0)}$ & $E_{(2,1,1)}$ & $E_{(2,2,1)}$ & $E_{(3,1,2)}$ & $E_{(3,2,2)}$ & $E_{(4,1,3)}$ & $E_{(4,2,3)}$ \\ \hline
0.0    & 3.73205  &  0.965425  &  0.968383  & 0.968026   &  0.968434  & 0.9682   & 0.968438   &  0.968285  & 0.96844   \\ \hline
0.2    & 3.72010  & 0.965226   &  0.968211  &  0.967851  &  0.968263  &  0.968051  &  0.968267  &  0.968112  & 0.968269   \\ \hline
0.4    & 3.68356  &  0.964601  &  0.967675  &  0.967301  & 0.967729   &  0.967509  & 0.967734   & 0.967572   &  0.967735  \\ \hline
0.6    & 3.62021  &  0.963458  & 0.966704   & 0.966304   & 0.966763   &  0.966526  &  0.966768  &  0.966593  & 0.96677  \\ \hline
0.8    & 3.52549  &  0.961588  & 0.965146   &  0.964698  &  0.965215  & 0.964945   &  0.96522  &  0.965021  &  0.965222 \\ \hline
1.0    & 3.39032  &  0.946508  &  0.962667  & 0.946508   &   0.962755 &  0.962419  &  0.962763  &  0.962512  & 0.962765  \\ 
\hline
\hline
\end{tabular}
}
\caption{The values of massive particle's energy for different configurations of periodic orbits characterized by $(z,w,v)$. Here, we set $L=\frac{1}{2}(L_{MBO}+L_{ISCO})$ and $\gamma=0.5$.}
\label{table1}
\end{table*}

In this section, we explore the periodic orbits around the dyonic ModMax BH. It should be noted that every periodic orbit is described by three integers, which are the zoom, whirl, and vertex numbers. One can write the ratio of the fundamental frequencies, which is called the rational number, in the following form~\cite{Levin_2008}
\begin{equation}
q=\frac{\omega_{\phi}}{\omega_{r}}-1=w+\frac{v}{z} ,
\end{equation}
with $\omega_r$ and $\omega_{\phi}$ refer to the radial and angular frequencies, respectively. We can rewrite the expression for the rational number by considering the equations of motion as follows~\cite{2025JCAP...01..091Y, SHABBIR2025101816,Jiang:2024cpe,Alloqulov:2025ucf}
\begin{eqnarray}
q&=&\frac{1}{\pi} \int_{r_1}^{r_2}\frac{\dot{\phi}}{\dot{r}}-1= \\ \nonumber
&=& \frac{1}{\pi}\int_{r_1}^{r_2} \frac{1}{r^2\sqrt{E^2-\left(1-\frac{2M}{r}+\frac{Q^2}{r^2}e^{-\gamma}\right)\left(1+\frac{L^{2}}{r^{2}}\right)}}
\end{eqnarray}
where $r_1$ and $r_2$ are the radii of periapsis and apoapsis of the periodic orbits, respectively. Fig.~\ref{q} represents the dependence of the rational number $q$ on the energy and orbital angular momentum for different values of the spacetime parameters. The figure shows that $q$ increases with increasing energy $E$, rising sharply as $E$ approaches its maximum.  In contrast, $q$ goes down with the increase of the energy $L$, rising sharply as $L$ approaches its minimum. Furthermore, the values of the rational number $q$ shift toward the lower side of the energy and orbital angular momentum under the influence of the BH charge. Additionally, we perform numerical computation to determine the values of the energy $E$ of periodic orbits which are characterized by $(z,w,v)$, while keeping orbital angular momentum as $L=\frac{1}{2}(L_{MBO}+L_{ISCO})$ in Table~\ref{table1}. Based on the results tabulated in Table~\ref{table1}, in Fig.~\ref{fig:periodicL}, we demonstrate the periodic orbits around the dyonic ModMax BH for various $(z,w,v)$ values using the bisection method for numerically solving complex equations. Here, we set the spacetime parameters as $Q=0.4$ and $\gamma=0.5$. As can be seen from Fig.~\ref{fig:periodicL}, massive particles with a greater zoom number $z$ exhibit periodic orbits with more complex structural patterns, whereas those with a higher whirl number $w$ make additional loops around the BH before reaching the next apoapsis.

\section{Numerical kludge gravitational waveforms from periodic orbits}
\label{numerical}

In this part, we explore the gravitational waveforms from periodic orbits around the dyonic ModMax BH. It should be emphasized that the extreme mass-ratio inspiral (EMRI) system is considered. The EMRI system involves a stellar-mass object along the periodic orbit around a supermassive dyonic ModMax BH. The GWs emitted by this EMRI system could contain imprints of the periodic orbits and the supermassive dyonic ModMax BH. One can use the widely recognized method for calculating the gravitational waveforms, which is the adiabatic approximation method, from an EMRI system. In the short term, both the energy and orbital angular momentum remain nearly constant, as the small object's orbital characteristics evolve over a time scale much longer than its orbital cycles. Thus, over a few orbital cycles, the trajectory can be approximated as a geodesic, and we define the gravitational waveforms produced by these periodic orbits over a complete cycle. It is worth noting that the effects of the gravitational radiation on the motion of the small object can be neglected for the given brief time frame. Using the numerical kludge waveform model, we explore the gravitational waveforms from the EMRI system. First, we compute numerical solutions to the equations of motion (given in Eq.~(\ref{eq:eqmotion})) to study periodic orbits of the small object orbiting a supermassive dyonic ModMax BH. Next, the gravitational waveforms can be generated by applying the symmetric and trace-free (STF) mass quadrupole equation for gravitational radiation~\cite{2025JCAP...01..091Y}, which can be written in the following form~\cite{RevModPhys.52.299}
\begin{equation}
I^{ij}=\Big[\int d^3 xx^ix^jT^{tt}(t,x^i)\Big]^{(\text{STF})}    
\end{equation}
where $T^{tt}$ refers to the $tt$-component of the stress-energy tensor for the small celestial body on the trajectory $Z^{i}(t)$. We can write its expression in the following form
\begin{equation}
T^{tt}(t,x^i)=m\delta^3(x^i-Z^i(t))\, .
\end{equation}
One can project the trajectory of the small celestial body into the Cartesian coordinate system by considering the Boyer-Lindquist coordinates as a fictitious spherical polar coordinate system
\begin{align}
x=r\sin{\theta}\cos{\phi}, \quad y=r\sin{\theta}\sin{\phi}, \quad z=r\cos{\theta}\, .
\end{align}
Metric perturbations can be obtained using the above equations in the following form
\begin{equation}
h_{ij}=\frac{2}{D_L}\frac{d^2I_{ij}}{dt^2}=\frac{2m}{D_L}(a_ix_j+a_jx_i+2v_iv_j),
\end{equation}
where $D_{L}$ refers to the luminosity distance from the detector to the EMRI system. $v_i$ and $a_i$ are the velocity and acceleration of the small celestial body, respectively. This waveform demonstrates the basic propagation characteristics of GWs in space. To match the real detectors, the modeled gravitational wave signal should be adjusted to align with the detector's framework. Therefore, we need to project the gravitational wave onto a coordinate system compatible with the detector~\cite{Poisson_Will_2014,2025JCAP...01..091Y,2025EPJC...85...36Z,2024arXiv241101858M}. One can write the coordinate basis as follows
\begin{equation}\label{coordsystem}
\begin{array}{l}
e_{X}=[\cos\zeta,-\sin\zeta,0]\,,\\
e_{Y}=[\cos\iota\sin\zeta,\cos\iota\cos\zeta,-\sin\iota]\,,\\
e_{Z}=[\sin\iota\sin\zeta,\sin\iota\cos\zeta,\cos\iota]\,.
\end{array}
\end{equation}
where $\zeta$ and $\iota$ refer to the longitude of the periastron and the angle of inclination of the orbit of the test particle relative to the observation direction, respectively. The vectors $e_X$, $e_Y$, and $e_Z$ represent the orthogonal coordinate basis of the detector's reference frame, which are used to break down the gravitational wave signal into its separate polarization components. Subsequently, the corresponding GW polarizations can be written in the following form~\cite{2025JCAP...01..091Y}
\begin{eqnarray}
 h_{+}&=&\frac{1}{2}(e^i_X e^j_X-e^i_Ye^j_Y)h_{ij}, \\
 h_{\times}&=&\frac{1}{2}(e^i_Xe^j_Y+e^i_Ye^j_X)h_{ij}.
\end{eqnarray}
Using Eq.~(\ref{coordsystem}), we can write the above equation as follows
\begin{eqnarray}
h_{+}&=&\frac{1}{2}(h_{\zeta\zeta}-h_{\iota\iota}), \\
h_{\times}&=& h_{\iota\zeta}.
\end{eqnarray}
where
\begin{eqnarray}
h_{\zeta\zeta}&=&h_{xx}\cos^2{\zeta}-h_{xy}\sin{2\zeta}+h_{yy}\sin^2{\zeta}, \\
h_{\iota\iota}&=&\cos{\iota}[h_{xx}\sin^2{\zeta}+h_{xy}\sin{2\zeta}+h_{yy}\cos^2{\zeta}]\nonumber\\&+&h_{zz}\sin^2{\iota}-\sin{2\iota}[h_{xz}\sin{\zeta}+h_{yz}\cos{\zeta}], \\
h_{\iota\zeta}&=&\cos{\iota}\Big[\frac{1}{2}h_{xx}\sin{2\zeta}+h_{xy}\cos{2\zeta}-\frac{1}{2}h_{yy}\sin{2\zeta}\Big]\nonumber\\&+&\sin{\iota}[h_{yz}\sin{\zeta}-h_{xz}\cos{\zeta}].
\end{eqnarray}
%
The EMRI system is considered to visualize the GWs. We set the parameters of the EMRI system as follows: the mass of the SMBH and the small object are $M \sim 10^7 M_{\odot}$ and $m \sim 10 M_{\odot}$, respectively. The latitude and angle of inclination $\zeta=\iota=\pi/4$, and the distance between light and dark $D_L=200\text{Mpc}$ are also set. Fig.~\ref{fig:combination} represents the gravitational waveforms radiated by the $(3,1,2)$ periodic orbit. One can see from this figure that the gravitational waveforms consist of the zoom and whirl stages. The calm part corresponds to the zoom stage, GW is radiated as the calm part when the small object enters the elliptical orbit far from the SMBH. In contrast, the rapid oscillation part refers to the whirl stage, and it is formed when the small object starts a whirl motion near the SMBH. Note that the dramatic rise in the gravitational wave frequency during the whirl stage is the direct cause of the intense oscillations. This figure also demonstrates the comparison results for the $Q=0.4$ and $Q=0.8$ cases, which are represented by blue and red lines, respectively. It can be seen that the zoom and whirl areas shrink with increasing charge, and this causes the GW signal to arrive earlier.  Furthermore, we analyze the parts of the periodic orbit (1,2,0) in Fig.~\ref{fig:combination2}. In other words, this figure gives the information about which part of the periodic orbits corresponds to which part of the gravitational waveforms. To be more informative, we make a comparison of the different periodic orbits in Fig.~\ref{fig:GW}. Our analysis is limited to a single orbital period, meaning that each orbit produces a distinct gravitational waveform.

\section{Conclusions}\label{conclusion}

In this work, our investigation begins with a brief analysis of the metric function of the dyonic ModMax BH, which allows us to understand how the structure of the event horizon is affected by the space-time parameters, namely, the BH charge $Q$ and screening parameter $\gamma$. We then demonstrated the regions in the $(Q,r)$ plane where the dyonic ModMax BH exists. Furthermore, we explored the motion of a massive particle around the dyonic ModMax BH using the radial equation of motion. Our findings showed that the effective potential increases with spacetime parameters as well as orbital angular momentum. We then determined the MBOs and ISCOs using the appropriate conditions. It was shown that spacetime parameters lead to higher values of both the MBO and ISCO. For further clarity, we presented Fig.~\ref{fig:EL}, the allowed parameter space of the energy and orbital angular momentum corresponding to bound orbits around the dyonic ModMax BH. We showed that increasing the BH charge shifts the allowed region toward smaller values of both energy and angular momentum for a massive particle, causing the periodic orbits to occur closer to the dyonic ModMax BH. In contrast, increasing the screening parameter shifts the allowed region toward larger values of both energy and angular momentum.         

Further, we examined the dependence of the rational number $q$ on the energy and orbital angular momentum for different values of the BH charge (see Fig.~\ref{q}). The results indicated that the rational number $q$ increases with increasing energy, while it decreases with increasing orbital angular momentum. In addition, the curves shift toward lower values of both energy and angular momentum as the BH charge grows. We then analyzed the periodic orbits characterized by the three integers ($z$,$w$,$v$). Using the numerical bisection method, we computed the corresponding energy values for various configurations of periodic orbits at $L=\frac{1}{2}(L_{MBO}+L_{ISCO})$, as listed in Table~\ref{table1}. Based on these results, we presented the associated periodic orbits, showing that massive particles' periodic orbits with a larger zoom number can be characterized by a more complex structure. In contrast, massive particles with a higher whirl number exhibit additional loops in the periodic orbits before reaching the subsequent apoapsis (see Fig.~\ref{fig:periodicL}).
 
Finally, we investigated the gravitational waveforms generated from periodic orbits around the dyonic ModMax BH. For this purpose, we considered an extreme mass-ratio inspiral (EMRI) system composed of a stellar-mass object ($m\sim10M_{\odot}$) orbiting a SMBH ($M\sim 10^7 M_{\odot}$). We showed the gravitational waveforms corresponding to the periodic orbit (3, 1, 2), computed using the numerical kludge method (see Fig.~\ref{fig:combination}). We observed that the zoom and whirl areas shrink with increasing the BH charge, which in turn causes the GW signal to arrive earlier. For a more comprehensive perspective, we also presented the periodic orbit (1,2,0) together with its associated gravitational waveforms (see Fig.~\ref{fig:combination2}), showing which parts of the periodic orbit correspond to the specific portions of the emitted gravitational waveform. Furthermore, we compared the gravitational waveforms generated from different periodic orbits (Fig.~\ref{fig:GW}). 

These theoretical results deepen our understanding of the underlying spacetime structure and provide valuable insights into the gravitational waveforms emitted by periodic orbits around the dyonic ModMax BH. Furthermore, analysis of these waveforms may offer an alternative approach for constraining the effects of the BH charge $Q$ and the screening parameter $\gamma$ in EMRI systems, particularly with observations from future new generation GW detectors.     

\section*{Acknowledgment}

S.S. is supported by the National Natural Science Foundation of China under Grant No. W2433018. T.Z. is supported by the National Natural Science Foundation of China under Grants No. 12275238 and 11675143, the National Key Research and Development Program under Grant No. 2020YFC2201503, and the Zhejiang Provincial Natural Science Foundation of China under Grants No. LR21A050001 and No. LY20A050002, and the Fundamental Research Funds for the Provincial Universities of Zhejiang in China under Grant No. RF-A2019015.

\bibliographystyle{apsrev4-1}
\bibliography{reference}
\end{document}